\documentclass{iopjournal}
\usepackage{graphicx}
\usepackage{url}
\usepackage[numbers,sort&compress]{natbib}
\def\eion{{(e~+~ion)}\ }

\def\eg{{\it{e.g.}}}

\def\ovii{{\rm O~\scalebox{0.8}{VII}}\ }
\def\ovi{{\rm O~\scalebox{0.8}{VI}}\ }
\def\oviii{{\rm O~\scalebox{0.8}{VIII}}\ }

\def\en{{$n$\ }}
\def\el{{$l$\ }}
\usepackage{amsmath}
\newcommand{\be}{\begin{equation}}
\newcommand{\ee}{\end{equation}}
\usepackage{subcaption}
\usepackage{float}
\usepackage{multirow}

\begin{document}

\articletype{Paper} %	 e.g. Paper, Letter, Topical Review...

\title{R-Matrix calculations for opacities:~V. Temperature-density dependence of photoabsorption cross sections and opacity spectra of oxygen ions \ovi and \ovii.}

\author{D Chari$^{1}$, S N Nahar$^2$, A K}
Pradhan$^{2,3,*}$

\affil{$^1$Department of Physics, The Ohio State University, Ohio 43210, USA}

\affil{$^2$Department of Astronomy, The Ohio State University, Ohio 43210, USA}

\affil{$^3$Chemical Physics Program, The Ohio State University, Ohio 43210, USA}

\affil{$^*$Author to whom any correspondence should be addressed.}

\email{pradhan.1@osu.edu}

\keywords{Photoionization, Atomic Processes, Photoabsorption, R-matrix method, Plasma Opacity}

\begin{abstract}
We present R-matrix photoabsorption cross sections for Li-like oxygen O\,\scalebox{0.8}{VI} and He-like oxygen O\,\scalebox{0.8}{VII} and examine how plasma broadening modifies them as a function of temperature and density in astrophysical and laboratory high-energy-density (HED) plasma sources. All atomic systems are subject to plasma environment effects, and the propagation of radiation depends on photoabsorption via bound-bound transitions as spectral lines and autoionizing resonances in bound-free photoionization cross sections. We identify and illustrate low and high temperature-density limits for the onset of plasma broadening in O\,\scalebox{0.8}{VI} and O\,\scalebox{0.8}{VII}, demonstrating general features and methodology applicable across a broad range of plasma conditions. Calculations are presented along two representative isotherms corresponding to the solar base of convection zone (BCZ) at  $T = 1 \times 10^6$ and $2 \times 10^6$\,K, and electron densities $N_e = 10^{18}$--$10^{23}$\,cm$^{-3}$. Autoionizing resonances progressively dissolve into the continuum with increasing electron density at each isotherm, flattening and merging into the background cross sections at BCZ conditions, at much lower densities than bound-bound line features require. Illustrative examples are given for energy regions containing Rydberg resonance series, including large photoexcitation-of-core (PEC) resonances. Comparisons with previous Opacity Project results reveal that (i) the R-matrix photoionization cross sections cover a much higher energy range where a significantly richer spectrum of autoionizing resonances are present that are not included in OP, and (ii) the corresponding monochromatic opacities show significant quantitative differences. This work is generally applicable to modeling HED plasmas in astrophysics, and to the analysis of transmission spectra in laboratory experiments on inertial confinement fusion (ICF) devices.

\end{abstract}

\noindent\rule{\textwidth}{0.4pt}

\noindent {\small This is the author-prepared version of an open access article published in \textit{Journal of Physics B: Atomic, Molecular and Optical Physics} \textbf{59} 145003 (2026) under a Creative Commons Attribution 4.0 International (CC BY 4.0) licence. The Version of Record is available online at \url{https://doi.org/10.1088/1361-6455/ae8c98}.}

\noindent\rule{\textwidth}{0.4pt}

\section{Introduction}

\begin{figure}
 \centering
        \includegraphics[width=1.0\textwidth]{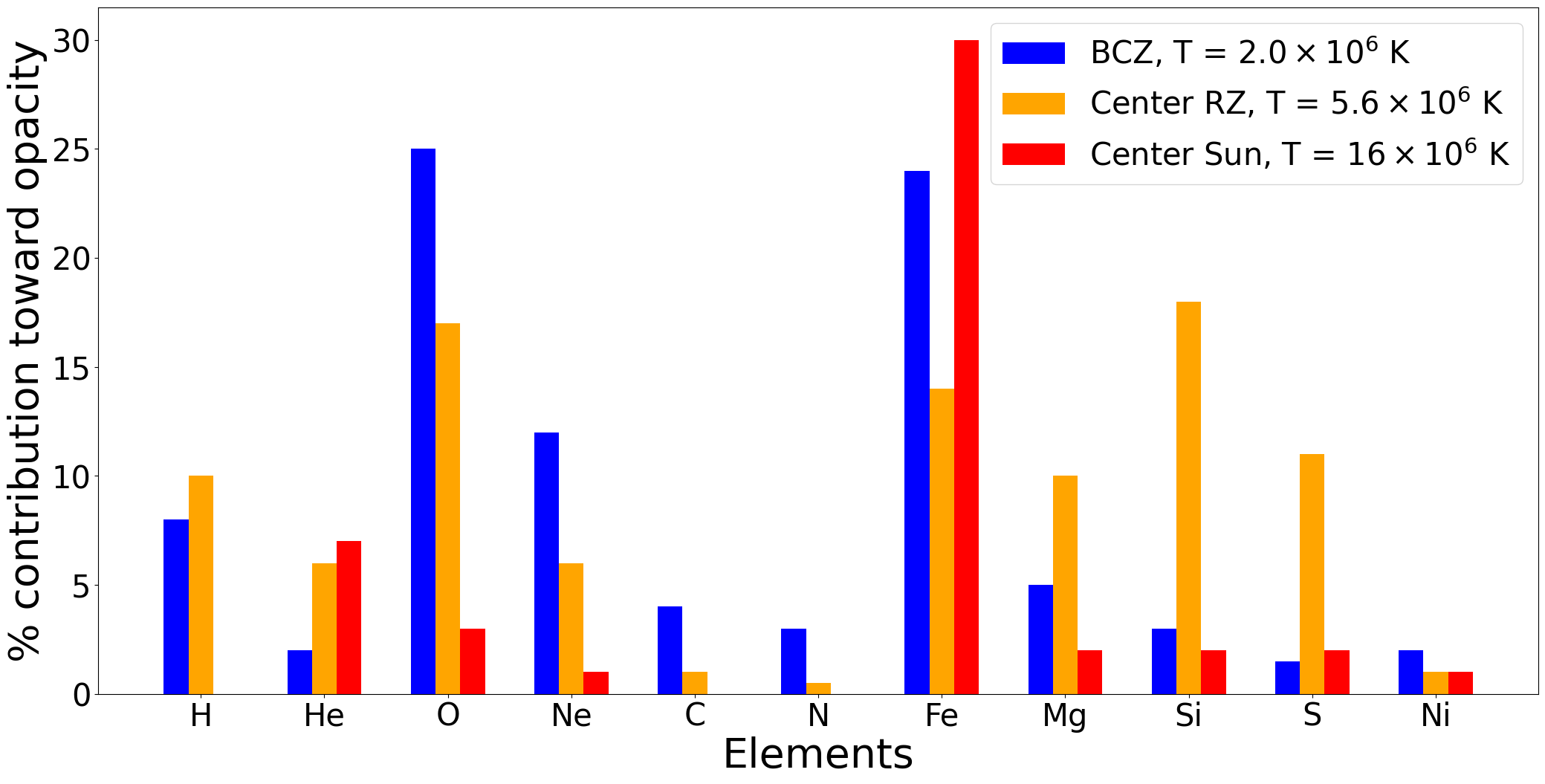}
 \caption{Opacity contributions of the most abundant elements in the solar interior at three depths: the base of the convection zone (BCZ, $T = 2.0 \times 10^6$\,K), the center of the radiative zone ($T = 5.6 \times 10^6$\,K), and the solar core ($T = 1.6 \times 10^7$\,K), computed using solar abundances from \cite{Asplund_2021}. Oxygen contributes $\sim$25\% of the total opacity at the BCZ, decreasing to 17\% in the radiative zone and 3\% in the core.} 
% Iron contributes $\sim$24\% at the BCZ, 14\% in the radiative zone, and increases to nearly 30\% at core temperatures.}

\label{fig1}
\end{figure}

Previous papers in this series broadly focus on large-scale R-matrix calculations of opacities (RMOP) for radiation transport models in high-energy-density (HED) plasma sources \cite{rmop1,p24}. Specific aims of the RMOP project are:  (1) to solve the long-standing solar abundances problem that originates from the substantially lower solar C, N, and O abundances obtained from modern 3D NLTE spectroscopic analyses \cite{Asplund_2021}. 
In contrast to previously higher solar elemental abundances, incorporating these reduced abundances into SSMs leads to significant discrepancies with helioseismic data, including larger deviations in the sound-speed profile, thermal structure, and an incorrect depth of the BCZ or the radiative-convection zone boundary, which is a precisely determined quantity from helioseismology to be 0.713 $\pm0.001$ R$_\odot$ \cite{buldgen25}. It is known that at the BCZ radiative opacity is dominated by oxygen and iron, which together account for nearly 50\% of the total and contribute equally. Figure~\ref{fig1} shows the percentage distribution of opacity from the most abundant elements in the Sun.

These considerations have motivated a fundamental reassessment of the 
atomic physics underlying opacity calculations. In previous RMOP papers I-IV \cite{rmop1,rmop2,rmop3,rmop4} we focused on iron ions with complex atomic structure. However, oxygen is 17 times more abundant than iron in the Sun, and although oxygen ions exist in highly or fully ionized states at the BCZ, their opacity also needs to be computed accurately, especially with HED plasma effects demonstrated in the RMOP papers. 

As shown in Table~\ref{tab1}, at the plasma temperatures and densities characteristic of the solar BCZ 
($T \approx 2.0 \times 10^6$\,K, $N_e \approx 10^{23}$\,cm$^{-3}$), oxygen ions are predominantly 
He-like \ovii and H-like \oviii, with \ovi present as a minor 
but non-negligible contributor. This is illustrated in Table~\ref{tab1}, which lists the 
dominant ionization fractions of oxygen at BCZ conditions. \ovi is more abundant at the lower isotherm ($T = 1 \times 10^6$\,K) and is included here as a representative Li-like ion.
The two ions of primary interest in this work—\ovi and \ovii—span the transition from 
Li-like to He-like configurations, and their opacity contributions are shaped by a rich 
interplay of resonance structures, electron-impact broadening, and configuration interaction. 
Theoretically, improved R-Matrix treatments of these effects point toward systematically 
higher opacities than previously predicted \cite{p24,pn25}. As shown in 
Figure~\ref{fig1}, oxygen contributes approximately 25\% of the total Rosseland mean opacity (defined below)
at the BCZ, making accurate atomic data for \ovi and \ovii essential to any resolution of 
the solar problem.

\begin{table}
\centering
\caption{Ionization states and fractions of Oxygen at BCZ - T=$2.0\times 10^{6}$ K and $N_{e} =10^{23}cm^{-3}$. Fractions below $10^{-4}$ are omitted. Remaining $\sim14\%$ distributed across lower ionization stages.}
\begin{tabular}{l c c }
\hline
Element & Ionization state & Ionization fraction at BCZ \\
\hline
Oxygen & O\,\scalebox{0.8}{VI} & $0.0002$ \\
& O\,\scalebox{0.8}{VII} &$ 0.12 $\\
& O\,\scalebox{0.8}{VIII} & $0.48$\\
 & O\,\scalebox{0.8}{IX} & $0.26$ \\

\hline
\end{tabular}
\label{tab1}
\end{table}

In this paper, following the treatment described in RMOP-I–III \cite{rmop1,rmop2,rmop3}, we present R-Matrix 
calculations of photoionization cross sections and monochromatic opacities for \ovi and 
\ovii, comparing our results with the earlier Opacity Project (hereafter OP) calculations. We further investigate
plasma broadening mechanisms across a broad range of temperatures and densities in order 
to map out plasma conditions under which broadening effects are significant for these ions. These ion-resolved results are building blocks for forthcoming total oxygen opacity and solar-mixture calculations.

%  Write a para for each item below
% - Mention the "solar problem" and the relationship between chemical abundances and opacities of constituent elements (Asplund \etal 2021, Buldgen \etal 2025, Pradhan and Nahar 2025).

% - Experimental measurements at Sandia Z (Bailey \etal 2015, Nagayama \etal 2019, and Livermore-Sandia-Los Alamos 2023).

% - Refer to RMOP-I table for Fe and Oxygen ion fractions at BCZ and Sandia Z conditions.

% - Fig. 1, a version of the bar chart of elemental opacity contributions, showing Fe and O opacity contributes nearly half and divided equally.

% - In this paper, following the treatment described in RMOP-I-III, we describe the atomic-plasma effects on individual Fe and O ions and calculate the variation and dependence of Rosseland Mean Opacity (reproduce expression).

\section{Theoretical framework}
 The R-matrix method \cite{pb11} and its adaptation to OP \cite{mjs87} are described in previous works, including the equation-of-state (EOS) for stellar interiors \cite{symp,aas11,p24,pn25}). Here, we reproduce and describe the basic expressions related to the physical quantities computed in this paper. Energies are reported in Rydbergs (1\,Ry = 13.6057\,eV) and in eV.
\subsection{Radiative processes and opacities}

Opacity is governed primarily by photon absorption via bound–bound
(photo–excitation) and bound–free (photoionization) transitions,
with free–free absorption (inverse bremsstrahlung) and photon scattering providing relatively smaller contributions. In a plasma source with a
mixture of elements with abundances $a_k$ and ionization fractions
$x_j$ of ionization stage $j$ of element $k$, the monochromatic
opacity can be written as
\begin{equation}
  \kappa_\nu = 
  \sum_k a_k \sum_j x_j
  \sum_{i,i'}
  \left[
    \kappa^{\rm bb}_{i i'}(\nu)
    + \kappa^{\rm bf}_{i \to \epsilon'}(\nu)
    + \kappa^{\rm ff}_{\epsilon \to \epsilon'}(\nu)
  \right]
  + \kappa^{\rm sc}_\nu,
\end{equation}
where $i,i'$ label bound levels, and $\epsilon,\epsilon'$ are
continuum electron energies. The different terms are expressed in
terms of oscillator strengths and photoionization cross sections
computed from atomic structure and scattering theory, and combined
with a plasma equation of state (EOS;\cite{p24}) that supplies level populations and ionization fractions.
In the present work we focus on (i) R-matrix photoionization cross sections \cite{snn98} instead of OP, and (ii) how plasma broadening of autoionizing (AI) resonances in bound–free cross sections modifies
$\kappa_\nu$ as a function of temperature $T(K)$ and electron
density $N_e(cm^{-3})$ for \ovi and \ovii.

\subsection{R–matrix atomic data for \ovi and \ovii}

The starting point of our calculation is a set of
unbroadened photoionization cross sections $\tilde{\sigma}_i(\omega)$
computed with the Breit–Pauli $R$–matrix (BPRM) method for \ovi
and \ovii \cite{snn98,norad}. In the close–coupling or coupled channel (CC) framework that underlies the R-matrix method \cite{pb11} the $(N+1)$–electron wavefunction of the $(e+{\rm ion})$ system in
a given symmetry is expanded as
\begin{equation}
  \Psi_E(e+{\rm ion}) =
  A \sum_i \chi_i({\rm ion}) \theta_i
  + \sum_j c_j \Phi_j ,
  \label{eq:cc}
\end{equation}
where $\chi_i$ are eigenfunctions of the $N$–electron target or core ion in states $S_i L_i (J_i) \pi_i$, $\theta_i$ describes the
incident electron in channels $i$, and the $\Phi_j$ are bound channel \eion functions that
account for short–range correlation and orthogonality.
Diagonalization of the Breit–Pauli Hamiltonian yields both
bound states ($E<0$) and continuum states ($E>0$), from which
oscillator strengths and photoionization cross sections are
obtained in a consistent way.

For \ovi (Li-like) and \ovii (He-like), we employ CC wavefunction expansions (Eq.~2) that include the relevant fine–structure levels of the next higher
ion as core states, up to  $n \leq 10$, $\ell \leq 9$ levels of the
$(e+{\rm ion})$ system. The CC wavefunction expansions include coupling between open ($E > 0)$ and closed ($E < 0$) channels in an ab initio manner. Quantum interference between open and closed channel wavefunctions generates dense series of
autoionizing resonances converging on to various target thresholds of the core ion, In particular, these include the strong Seaton resonances as photoexcitation–of–core (PEC) features, which are known to
strongly influence the bound–free opacity (\eg \cite{rmop2}). The computed relativisitic BPRM cross sections with fine structure (hereafter referred to as RMOP cross sections) are compared with available OP data computed in LS coupling to assess the role of additional correlation, relativistic effects, and resonances, included in the present work.

The unbroadened cross sections $\tilde{\sigma}_i(\omega)$ thus
include the intrinsic AI widths from channel coupling in the
RMOP cross sections, but do not yet include extrinsic plasma
broadening due to the ambient plasma environment.

\subsection{Plasma broadening of autoionizing resonances}

To model plasma effects on the intrinsic unbroadened resonance
profiles, we follow the general formulation developed in
RMOP~III \cite{rmop3,pn25}. In a dense HED plasma, autoionizing resonance profiles are attenuated and reshaped — in energy, width, height, and peak magnitude — by electron collisions, ion microfields (Stark effect), thermal Doppler broadening, and free-free transitions. We incorporate these effects, together with their corresponding partial widths, through a Lorentzian convolution of the unbroadened cross sections.

For a given bound initial level $i$, the plasma–broadened
photoionization cross section at photon energy $\omega$ is
\begin{equation}
  \sigma_i(\omega; T, N_e) =
  \int_{-\infty}^{+\infty}
  \tilde{\sigma}_i(\omega')\,
  \varphi(\omega',\omega; T, N_e)\,
  {\rm d}\omega' ,
  \label{eq:sigma_conv}
\end{equation}
where $\tilde{\sigma}_i(\omega')$ is the CC RMOP cross section with
fully resolved AI resonances, and $\varphi$ is a normalized
profile factor. We adopt a Lorentzian form
\begin{equation}
  \varphi(\omega',\omega; T, N_e) =
  \frac{\Gamma(\omega; T, N_e)/\pi}
       {(\omega-\omega')^2 + \Gamma^2(\omega; T, N_e)},
  \label{eq:lorentz}
\end{equation}
with the total width given by
\begin{equation}
  \Gamma(\omega; T, N_e) =
  \Gamma_{\rm c}(\omega; T, N_e) +
  \Gamma_{\rm s}(\omega; T, N_e) +
  \Gamma_{\rm d}(\omega; T) +
  \Gamma_{\rm ff}(\omega; T, N_e).
  \label{eq:gamma_total}
\end{equation}
The individual terms represent collisional (electron–impact),
Stark (ion–microfield), Doppler (thermal), and free–free (inverse bremsstrahlung) contributions
respectively, and are parametrized following the line–broadening
analogy adapted in RMOP~III \cite{rmop3}. 
At the BCZ plasma conditions considered herein $(T \sim 10^6 K, N_e = 10^{22} - 10^{23} cm^{-3})$, electron-impact broadening dominates over Stark, Doppler, and free-free contributions by typically an order of magnitude in the resonance region, justifying the Lorentzian convolution as a good representation of the net profile.

A key aspect of the formulation is that autoionization resonances
naturally occur in Rydberg series converging to a given excited core threshold $E_i$ of the residual ion. It is therefore convenient to recast the energy variable in terms of an effective quantum number $\nu_i$ relative to each threshold, and compute the
broadening widths $\Gamma(\nu_i; T, N_e)$ in that representation
before transforming back to $\omega$ for the convolution.
This procedure preserves the intrinsically asymmetric shapes of
the resonances while allowing their redistribution, flattening,
and eventual dissolution into the continuum as $T$ and $N_e$
increase.

The output of this step is a set of broadened cross sections
$\sigma_i(\omega; T, N_e)$ for each bound level $i$, on the same
energy mesh as the original CC RMOP data, for every
$(T, N_e)$ pair in the chosen grid.

\section{Computational method}

The main computational steps and modules involved in the present RMOP calculations are as follows.

\subsection{Input data and plasma grid}

As input we use level-resolved BPRM photoionization cross sections 
$\tilde{\sigma}_i(\omega)$ for O~\scalebox{0.8}{VI} and O~\scalebox{0.8}{VII}, computed on a fine 
energy mesh sufficient to resolve all AI resonance structures relevant for opacity, 
along with corresponding oscillator strengths and level energies to construct 
bound--bound and bound--free monochromatic opacities, and an LTE plasma EOS \cite{symp,aas11}
that yields ionization fractions and level populations as functions of temperature and density $(T(K),\rho(g/cc))$. 
The temperature--density domain is chosen to span the plasma conditions at the BCZ, and is sampled on a regular grid in $\log T(K)$ and $\log N_e(cm^{-3})$. The grid spans $\log T = 6.0$ and $6.3$ (i.e., $T = 1 \times 10^6$\,K and $2 \times 10^6$\,K) 
and $\log N_e = 15$--$23$ in steps of $1.0$, altogether yielding $18$ $(T, N_e)$ points along the two isotherms.

\subsection{Plasma broadening}
The plasma broadening in RMOP calculations is implemented as a post-processing module acting on the CC BPRM cross sections. For each ion, level $i$, and $(T,N_e)$ pair, the algorithm proceeds 
as follows (adapted from RMOP~III): for each excited core threshold $E_k$ included 
in the CC expansion, we map the photon energy grid $\omega'$ to the effective quantum 
number $\nu_k(\omega')$ of the Rydberg series with autoionizing resonances converging to $E_k$. We then evaluate the 
partial widths $\Gamma_{\rm c}, \Gamma_{\rm s}, \Gamma_{\rm d}, \Gamma_{\rm ff}$ at 
$(\nu_k; T, N_e)$ using the analytic prescriptions of RMOP~III (electron-impact damping, Stark cutoffs, Doppler broadening, and free-free transitions), to form the total width FWHM
$\Gamma(\omega; T, N_e)$ via Eq.~\eqref{eq:gamma_total}. For each energy point 
$\omega$ on the original mesh, we compute the Lorentzian profile 
$\varphi(\omega',\omega; T, N_e)$ with an integration interval $[\omega-\Delta,
\omega+\Delta]$, where $\Delta = \Gamma/\sqrt{\delta}$ and $\delta$ is a 
convergence parameter ensuring that the full profile (including wings) is captured 
to the desired accuracy; usually $\delta = 0.01$ which implies that the Lorentzian profile at each energy extends to 10 $\times$ FWHM $\Gamma$. Finally, we perform the convolution integral in 
Eq.~\eqref{eq:sigma_conv} numerically to obtain $\sigma_i(\omega; T, N_e)$ on the 
same energy grid. The loops over $(T,N_e)$, core thresholds, and levels are 
structured so that the same input unbroadened BPRM cross sections are re-used and the 
plasma-dependent widths and convolutions are recomputed at each energy. Computational CPU time for a complete convolution over all BPRM cross sections for a given ion, at each temperature-density $(T,N_e)$ pair, scale mainly with density for an isotherm through increasing $\Gamma$ and corresponding integration range, and range from a few minutes to hours for the grids considered here.

\subsection{Monochromatic opacities}
Once the broadened cross sections are computed, we calculate the RMOP monochromatic and mean opacities at each $(T,N_e)$ on a common photon energy grid \cite{p24}. The bound-free opacity 
is obtained by replacing $\tilde{\sigma}_i(\omega)$ with $\sigma_i(\omega; T, N_e)$ 
in the usual bound-free opacity expression (Eq.~1), with level populations supplied by 
the Mihalas-Hummer-D\"{a}ppen EOS (MHD-EOS \cite{symp,p24}). Bound-bound line profiles are treated with the standard line-broadening 
formalism used in the earlier OP work \cite{symp},  with the same set of parameters for radiation damping, collisional and Stark broadening, free-free and photon scattering. This yields $\kappa_\nu(T,N_e)$ for \ovi and \ovii 
covering the full energy range across the temperature and density grid of interest. An important improvement over OP, in addition to relativistic fine structure as opposed to LS coupling, is that the radiation damping widths are computed using A-values for fine structure transitions among bound-bound levels computed in the BPRM approximation \cite{norad}. 
The resulting RMOP monochromatic opacities are therefore obtained consistently using BPRM data for bound-bound and bound-free transitions, and results are compared with OP calculations 
to assess the impact of the present BPRM data and plasma broadening.

%\newpage
\section{Results and Discussion}
 Present RMOP results for \ovi and \ovii are compared with the data and codes used to compute existing OP opacities in the database OPserver \cite{mendoza07}. Owing to the fact that OP results are in LS coupling, we can compare the present BPRM data for LS terms with only one fine structure level, i.e. with S = 0 for singlets (2S+1) = 1. In addition, there are two important differences, the OP energy range is quite small and there are no autoionizing resonances.
 
\subsection{Li-like \ovi}
 Both the OP and BPRM calculations for Li-like \ovi go up to energy levels with \en=10 and \el=9. We compare a sample that elucidate the main differences. The OP calculations for the lithium isoelectronic sequence ions were in LS coupling and included a semi-empirical model potential to construct He-like target ion eigenfunctions \cite{Peach_1988}.

\subsubsection{Photoionization cross sections }

\begin{figure}[H]
    \centering
    \begin{subfigure}{0.48\textwidth}
        \centering
        \includegraphics[width=\textwidth]{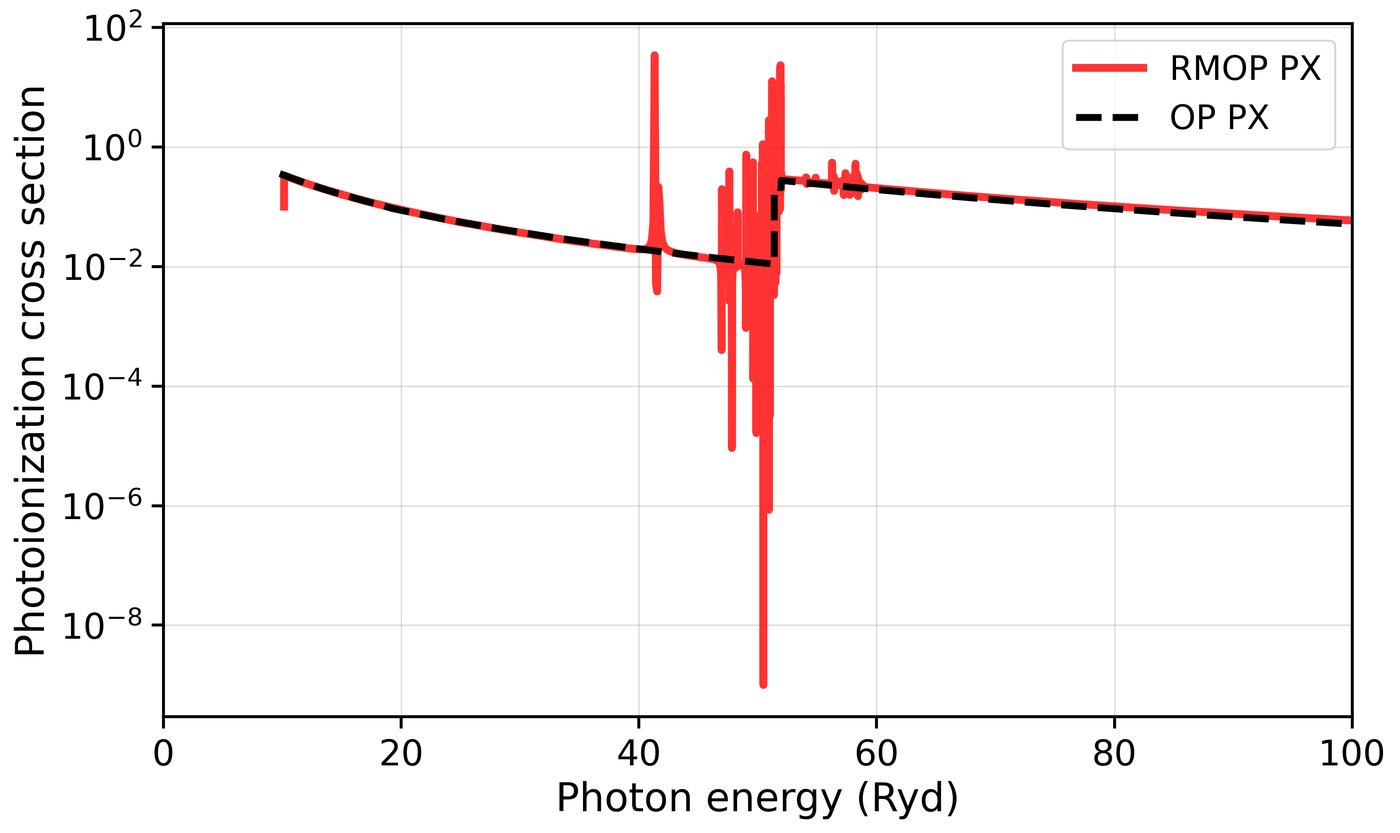}
        \caption{$1s^2\,(^{1}S)\,2s\,a^{2}S_{1/2}$, E = 10.1 Ry}
    \end{subfigure}
    \hfill
    \begin{subfigure}{0.48\textwidth}
        \centering
        \includegraphics[width=\textwidth]{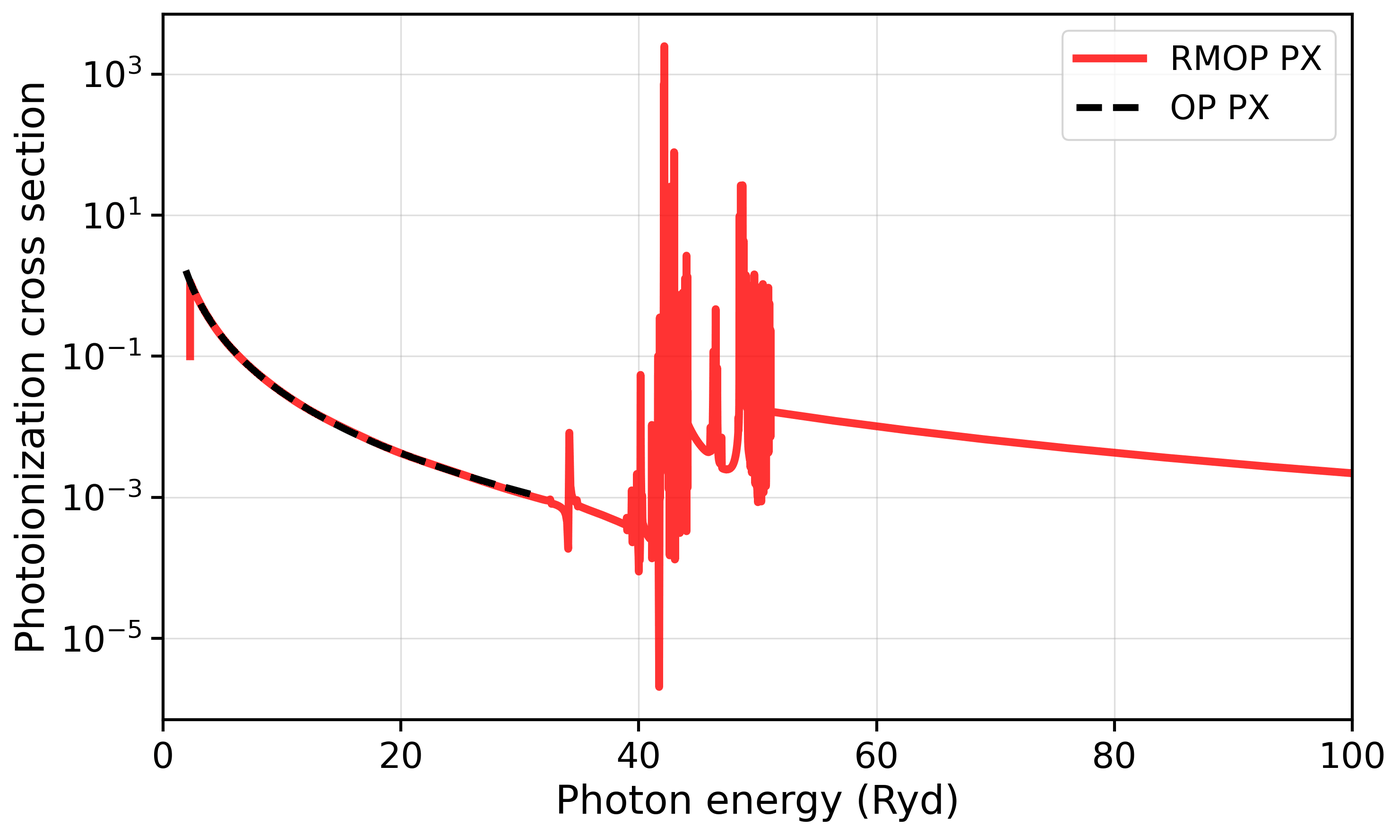}
        \caption{$1s^2\,(^{1}S)\,4p\,x^{2}P^{\circ}$, E = 2.28 Ry}
    \end{subfigure}

    \vspace{1em}

    \begin{subfigure}{0.48\textwidth}
        \centering
        \includegraphics[width=\textwidth]{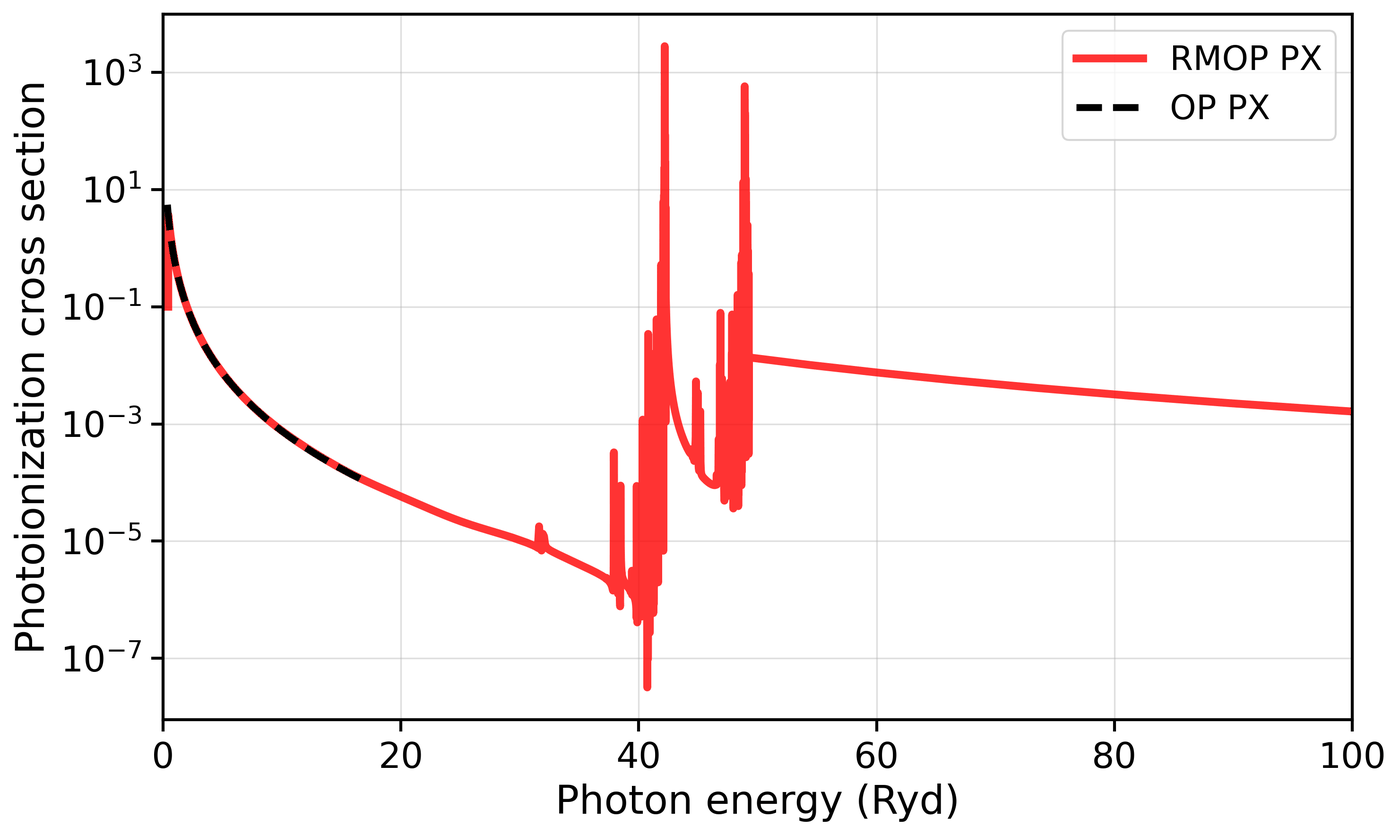}
        \caption{$1s^2\,(^{1}S)\,9d\,g^{2}D$, E = 0.44 Ry}
    \end{subfigure}
    
   \caption{Comparison of OP and RMOP photoionization cross sections for selected \ovi levels (out
of 26 total). The RMOP cross sections exhibit a richer resonance structure and extend over a broader
photon-energy range, while the OP data show no resonances and terminate at lower energies.}
    \label{fig:px_o6}
\end{figure}

Figure~\ref{fig:px_o6} shows BPRM photoionization cross sections for three representative 
levels and ionization energies of O~\scalebox{0.8}{VI}—$1s^2\,(^{1}S)\,2s\,a^{2}S_{1/2}$ (E = 10.1 Ry), 
$1s^2\,(^{1}S)\,4p\,x^{2}P^{\circ}$ (E = 2.28 Ry), and $1s^2\,(^{1}S)\,9d\,g^{2}D$ 
(E = 0.44 Ry) compared with OP data. The OP database TOPbase \cite{mendoza07} contains 26 LS terms against the 54 fine structure levels in the present RMOP calculations; identified and matching by corresponding ionization energies. 

Above the ionization threshold of Li-like \ovi $1s^2 \ ^1S_0$ in the non-resonant region, both the OP and BPRM photoionization cross sections decrease with energy and have the expected hydrogenic form. The next excited thresholds corresponds to He-like \ovii \en = 2 and \en=3 configurations that lie quite high in energy. Below those,the RMOP cross sections exhibit a rich autoionizing resonance structure arising from close-coupling channel interactions, with peaks spanning 
several orders of magnitude above the continuum, while the OP data show no resonances and 
terminate at much lower photon energies. The RMOP cross sections extend continuously to 
$\sim 250$~Ry, capturing the full bound-free continuum above all core thresholds of He-like \ovii (full extent not shown for clarity; see supplementary data in \cite{norad}).

The resonances form a Rydberg series converging to excited O~\scalebox{0.8}{VII} thresholds and 
become increasingly dense toward higher $n$. As level ionization energy decreases from the $2s$ to 
the $9d$ state, the threshold shifts to lower photon energies and the resonance structure 
concentrates in a narrower window, a trend consistent across all 24 matched levels. These 
missing resonances and the much lower energy extent in the OP data represent a significant shortcoming for opacity 
calculations under HED plasma conditions where this energy region dominates the bound-free opacity.

\subsubsection{Plasma broadening of photoionization cross sections}

\begin{figure}[H]
    \centering

    % Row 1
    \begin{subfigure}[b]{0.49\textwidth}
        \includegraphics[width=\textwidth]{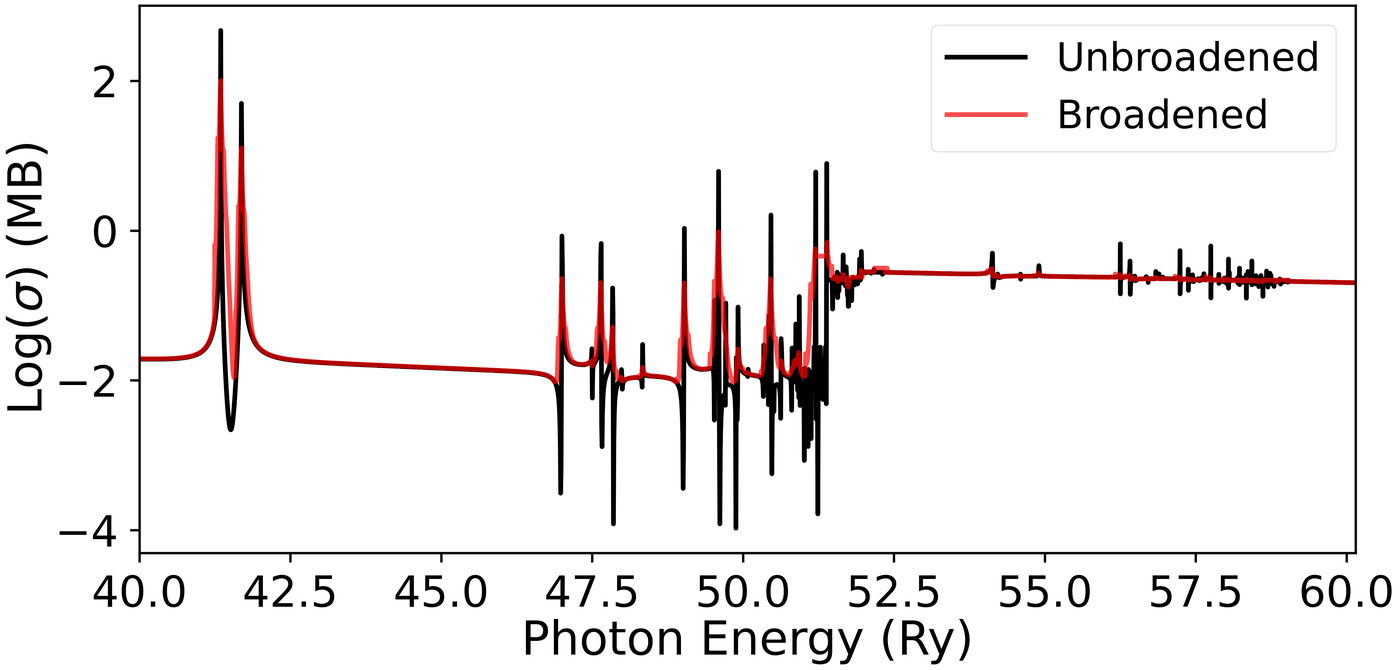}
        \caption{$N_e = 10^{18}$}
    \end{subfigure}
    \hfill
    \begin{subfigure}[b]{0.49\textwidth}
        \includegraphics[width=\textwidth]{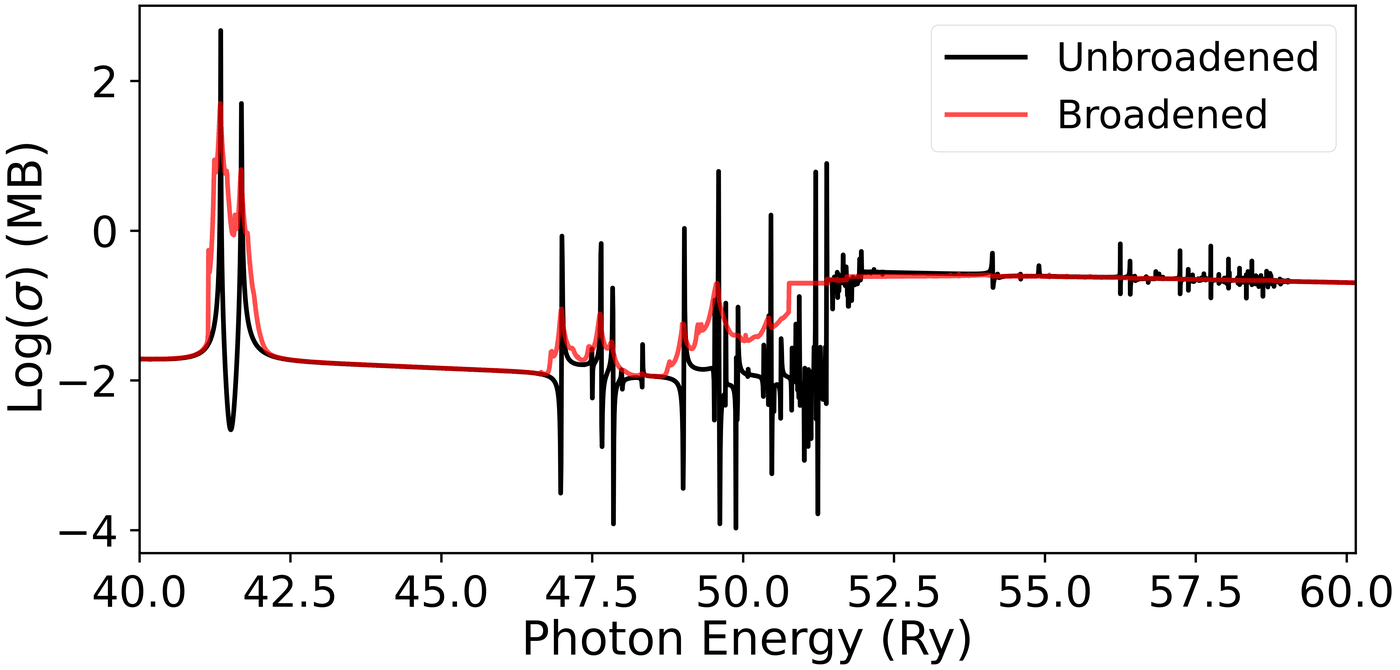}
        \caption{$N_e = 10^{20}$} 
    \end{subfigure}

    \vspace{1em}

    % Row 2
    \begin{subfigure}[b]{0.49\textwidth}
        \includegraphics[width=\textwidth]{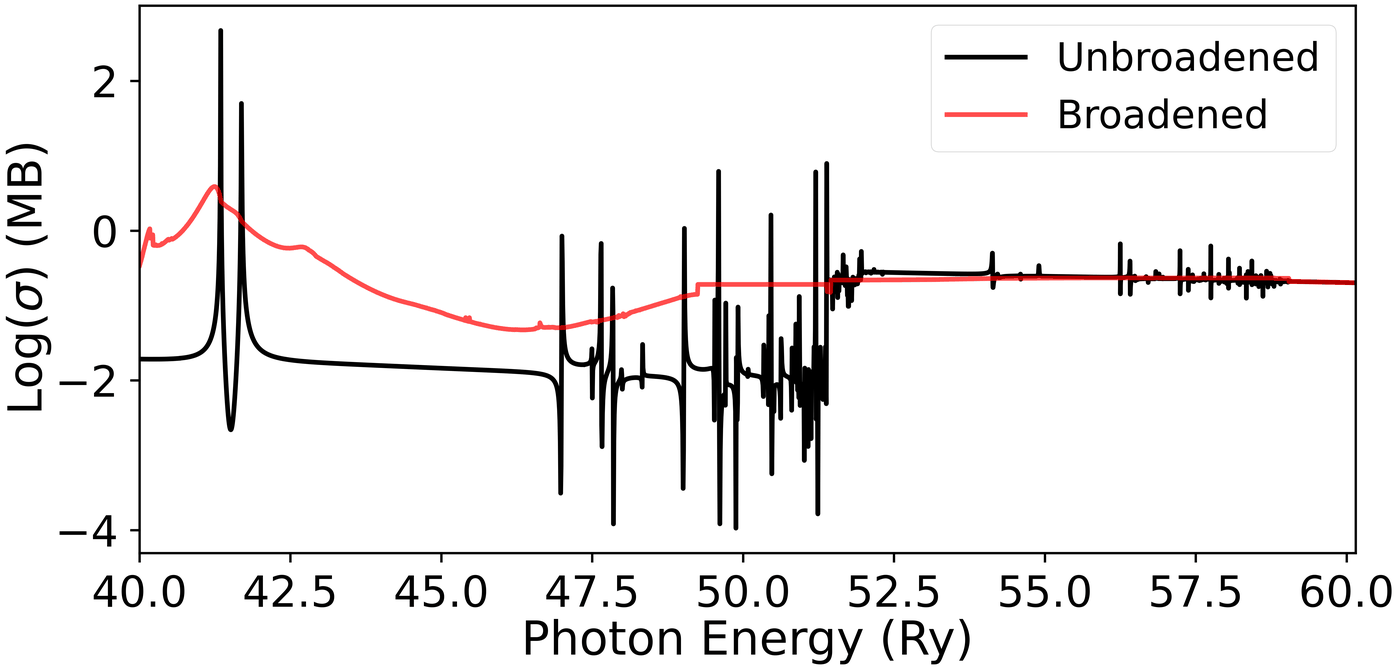}
        \caption{$N_e = 10^{22}$}
    \end{subfigure}
    \hfill
    \begin{subfigure}[b]{0.49\textwidth}
        \includegraphics[width=\textwidth]{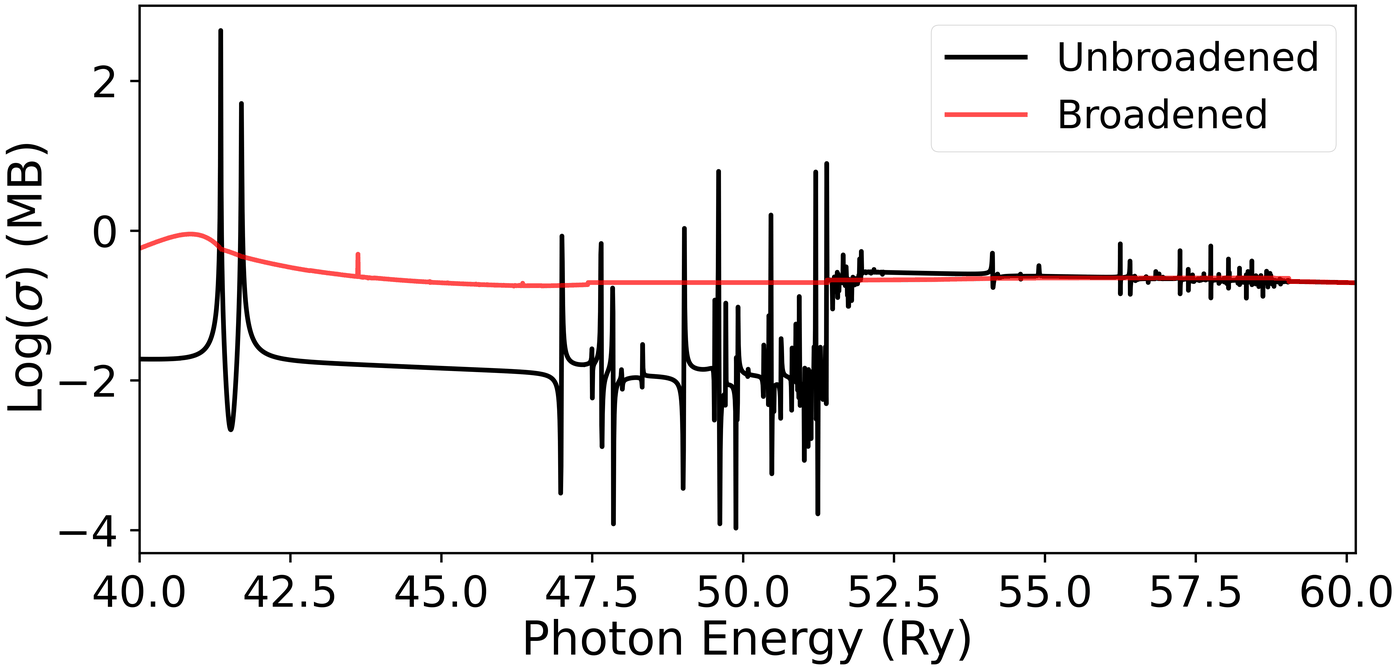}
        \caption{$N_e = 10^{23}$}
    \end{subfigure}

    \vspace{1em}

    \caption{Logarithmic plasma broadened photoionization cross sections for 
    $\hbar\omega + \mathrm{O\,VI} \rightarrow e + \mathrm{O\,VII}$ 
    from the bound level $1s^2\,(^{1}S)\,2s\,a^{2}S_{1/2}$ at 
    $\mathbf{T =1 \times 10^{6}\,\mathrm{K}}$ for varying electron densities. Towards the high densities and energies shown, the large unbroadened resonance structures dissolve into a flatter background, which is raised considerably due to plasma broadening effects.
    }
    \label{fig:log_1e6}
\end{figure}

\begin{figure}[H]
    \centering

    % Row 1
    \begin{subfigure}[b]{0.49\textwidth}
        \includegraphics[width=\textwidth]{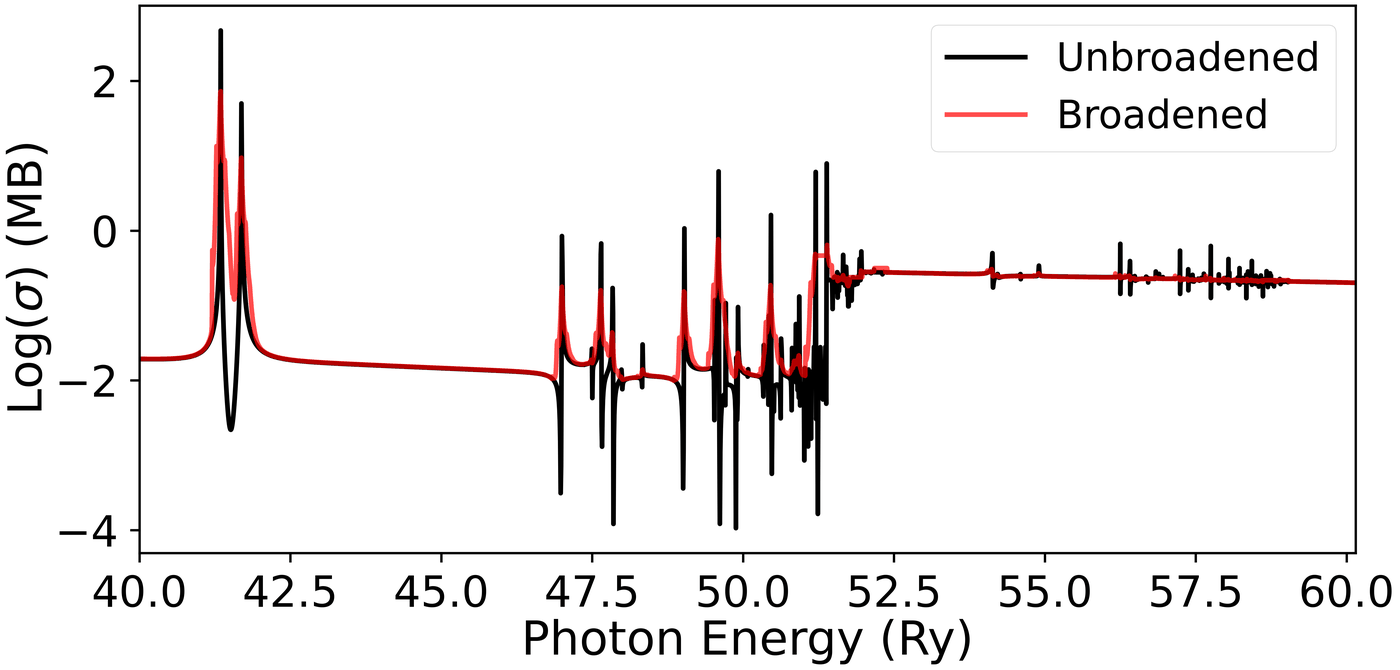}
        \caption{$N_e = 10^{18}$}
    \end{subfigure}
    \hfill
    \begin{subfigure}[b]{0.49\textwidth}
        \includegraphics[width=\textwidth]{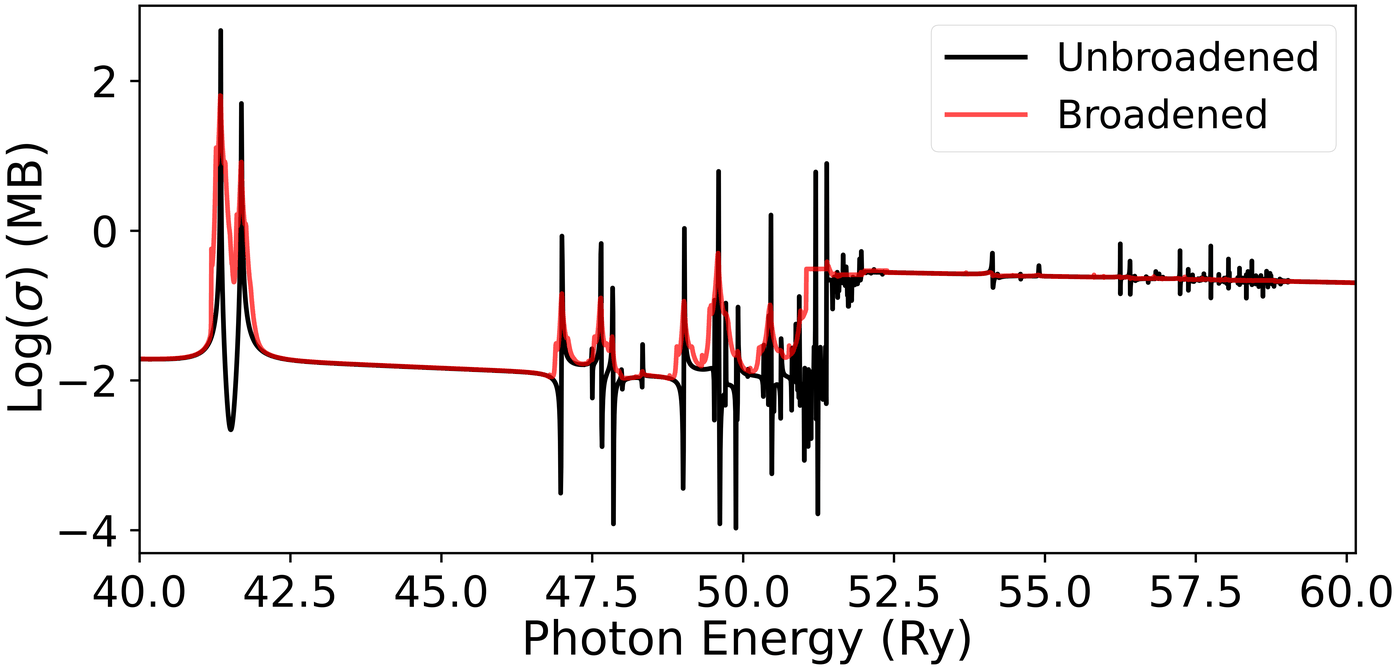}
        \caption{$N_e = 10^{19}$}
    \end{subfigure}

    \vspace{1em}

    % Row 2
    \begin{subfigure}[b]{0.49\textwidth}
        \includegraphics[width=\textwidth]{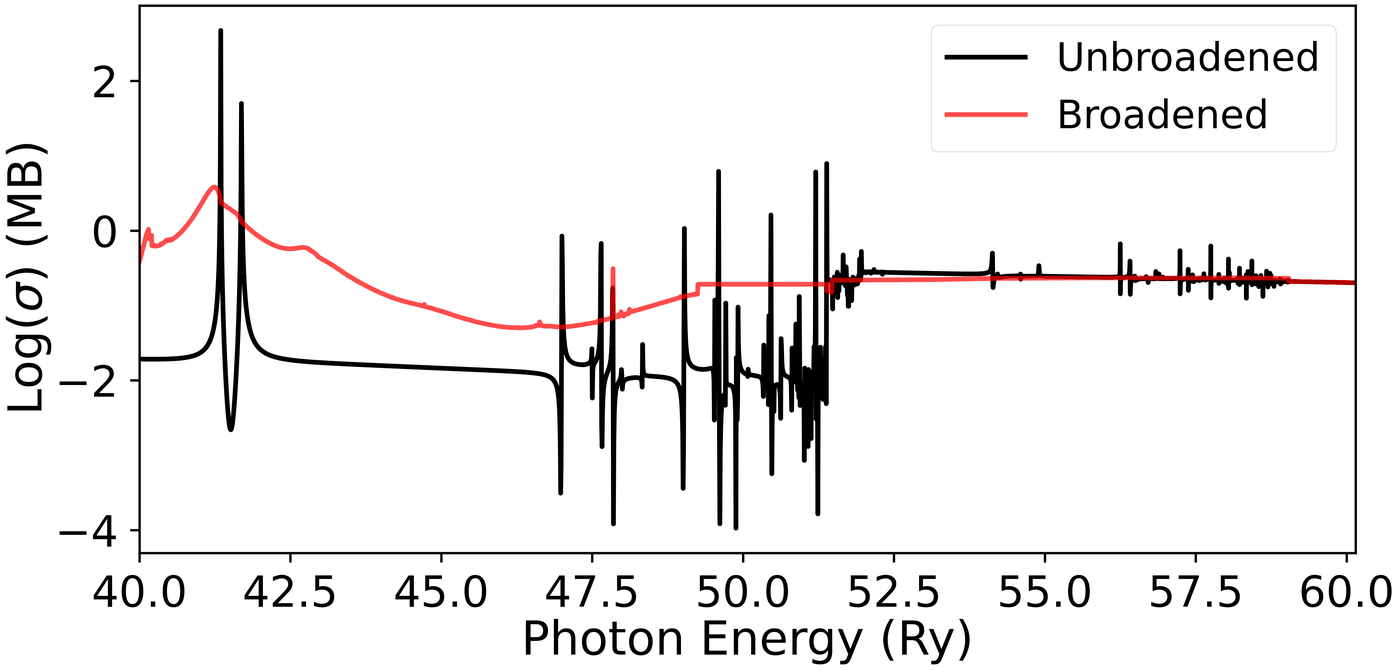}
        \caption{$N_e = 10^{22}$}
    \end{subfigure}
    \hfill
    \begin{subfigure}[b]{0.49\textwidth}
        \includegraphics[width=\textwidth]{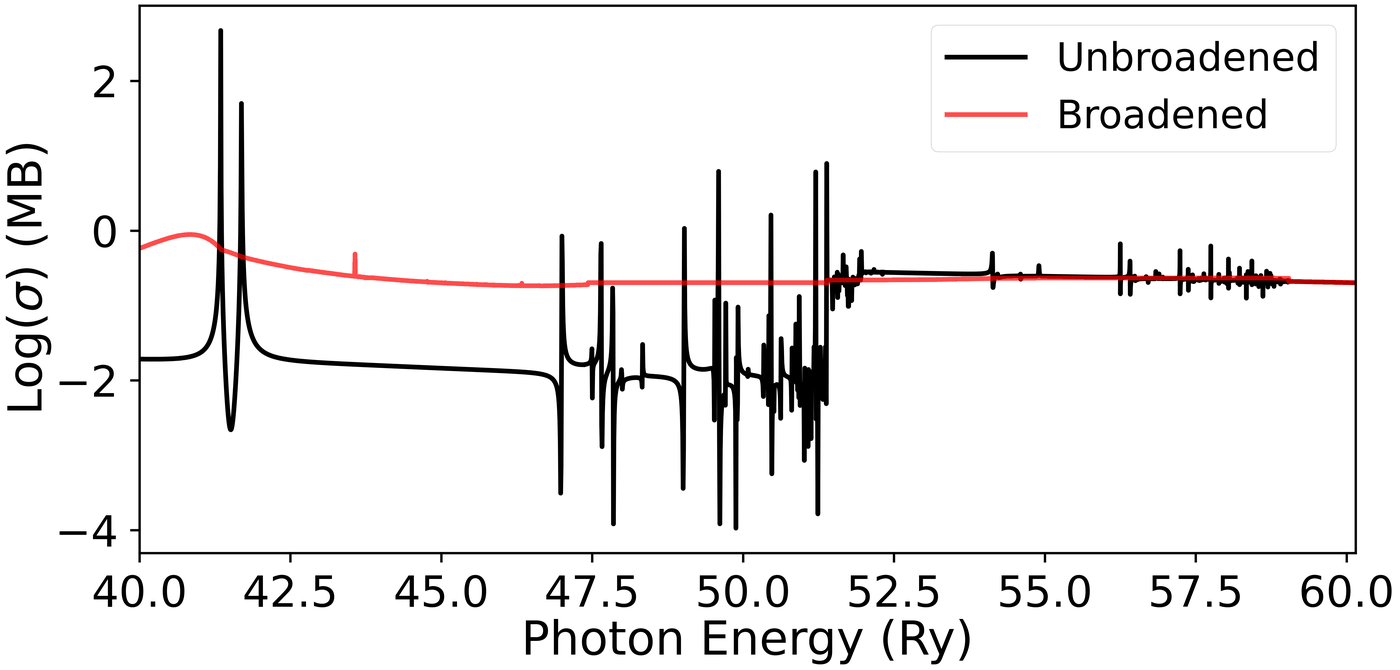}
        \caption{$N_e = 10^{23}$ BCZ}
    \end{subfigure}

    \vspace{1em}

    \caption{Logarithmic plasma broadened photoionization cross sections for 
    $\hbar\omega + \mathrm{O\,VI} \rightarrow e + \mathrm{O\,VII}$ 
    from the bound level $1s^2\,(^{1}S)\,2s\,a^{2}S_{1/2}$ at 
    $\mathbf{T =2 \times 10^{6}\,\mathrm{K}}$ for varying electron densities.
    }
    \label{fig:log_2e6}
\end{figure}

\begin{figure}[H]
    \centering

    % Row 1
    \begin{subfigure}[b]{0.49\textwidth}
        \includegraphics[width=\textwidth]{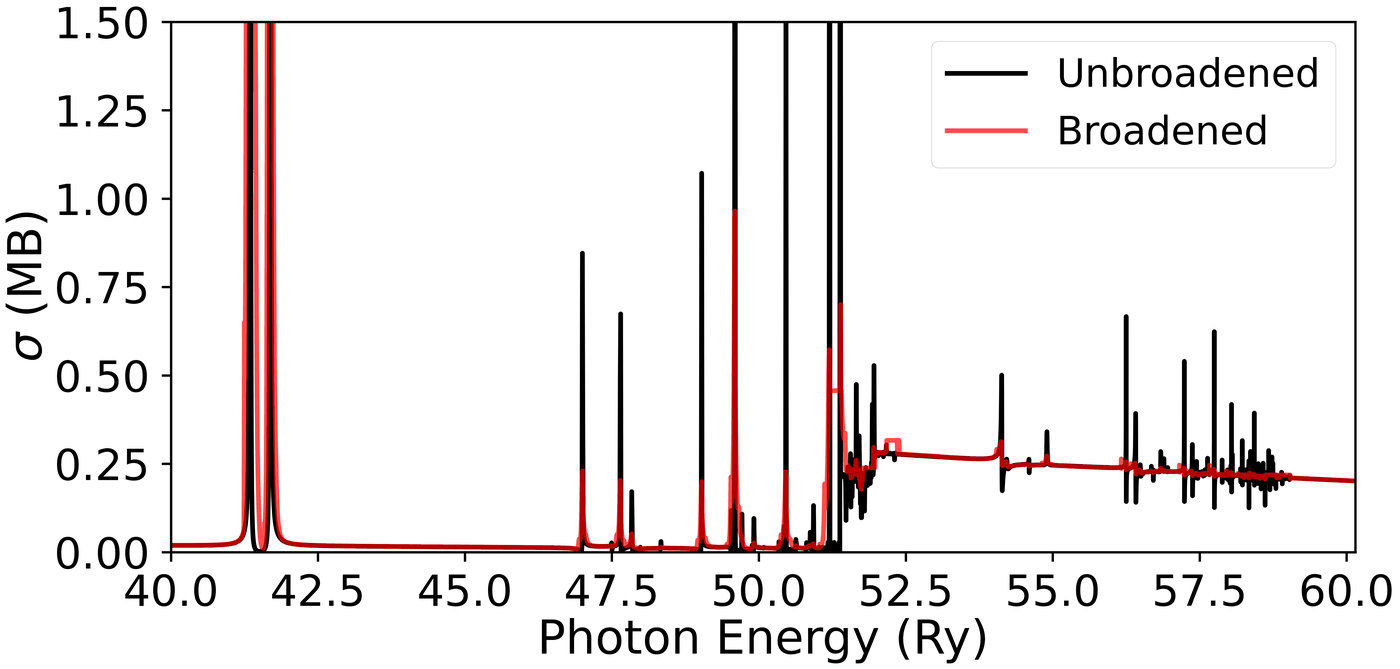}
        \caption{$N_e = 10^{18}$}
    \end{subfigure}
    \hfill
    \begin{subfigure}[b]{0.49\textwidth}
        \includegraphics[width=\textwidth]{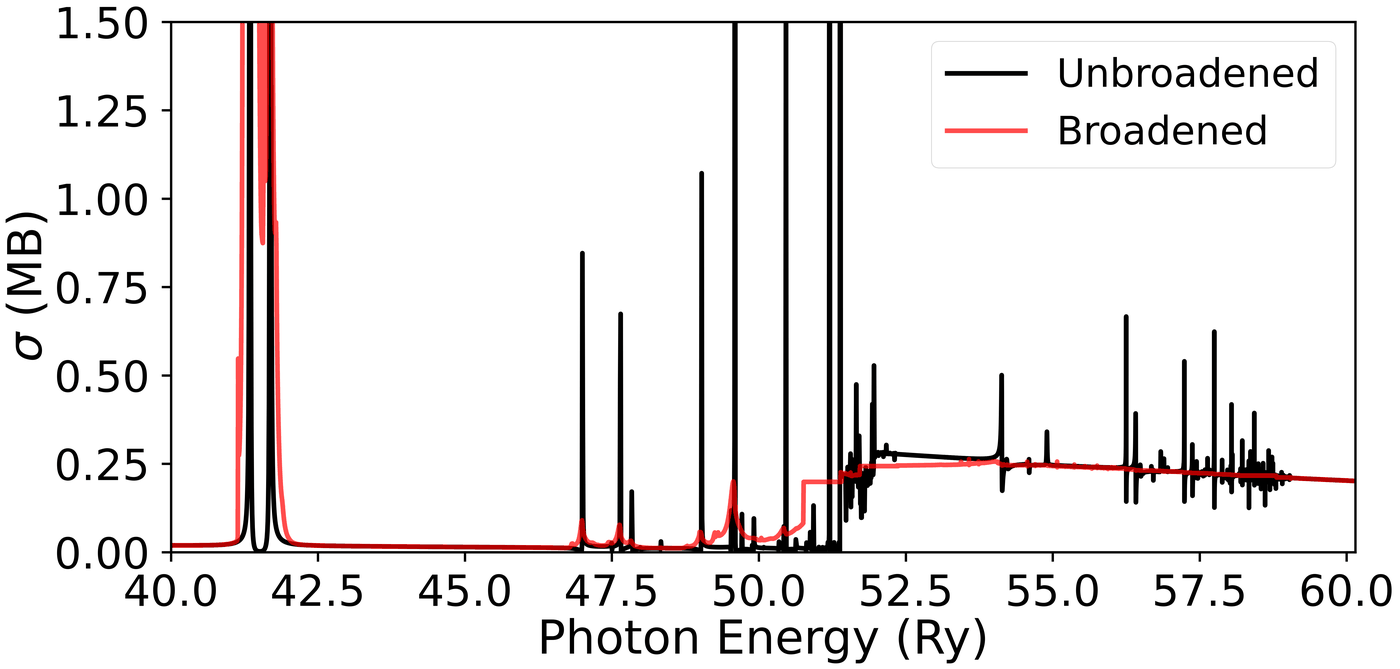}
        \caption{$N_e = 10^{20}$}
    \end{subfigure}

    \vspace{1em}

    % Row 2
    \begin{subfigure}[b]{0.49\textwidth}
        \includegraphics[width=\textwidth]{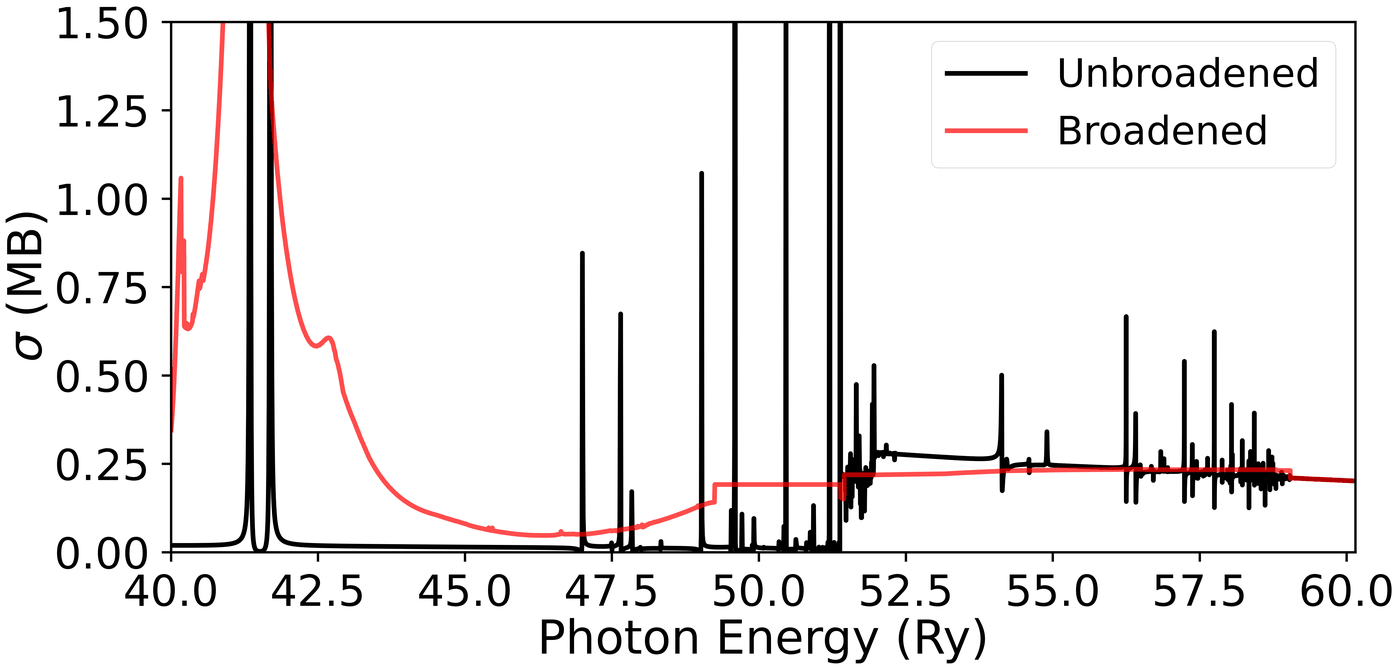}
        \caption{$N_e = 10^{22}$}
    \end{subfigure}
    \hfill
    \begin{subfigure}[b]{0.49\textwidth}
        \includegraphics[width=\textwidth]{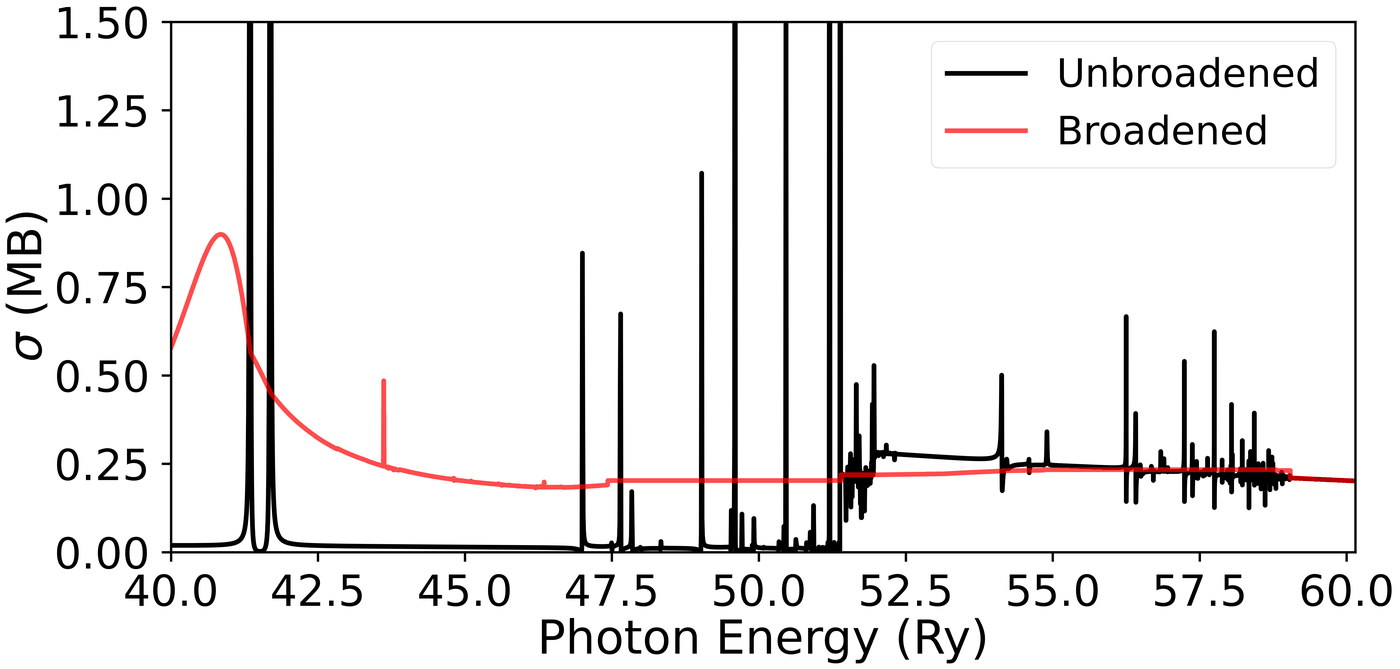}
        \caption{$N_e = 10^{23}$}
    \end{subfigure}

    \vspace{1em}

    \caption{Linear scale: plasma broadened photoionization cross sections for 
    $\hbar\omega + \mathrm{O\,VI} \rightarrow e + \mathrm{O\,VII}$ 
    from the bound level $1s^2\,(^{1}S)\,2s\,a^{2}S_{1/2}$ at 
    $\mathbf{T =1 \times 10^{6}\,\mathrm{K}}$ for varying electron densities.
    }
    \label{fig:lin_1e6}
\end{figure}

\begin{figure}[H]
    \centering

    % Row 1
    \begin{subfigure}[b]{0.49\textwidth}
        \includegraphics[width=\textwidth]{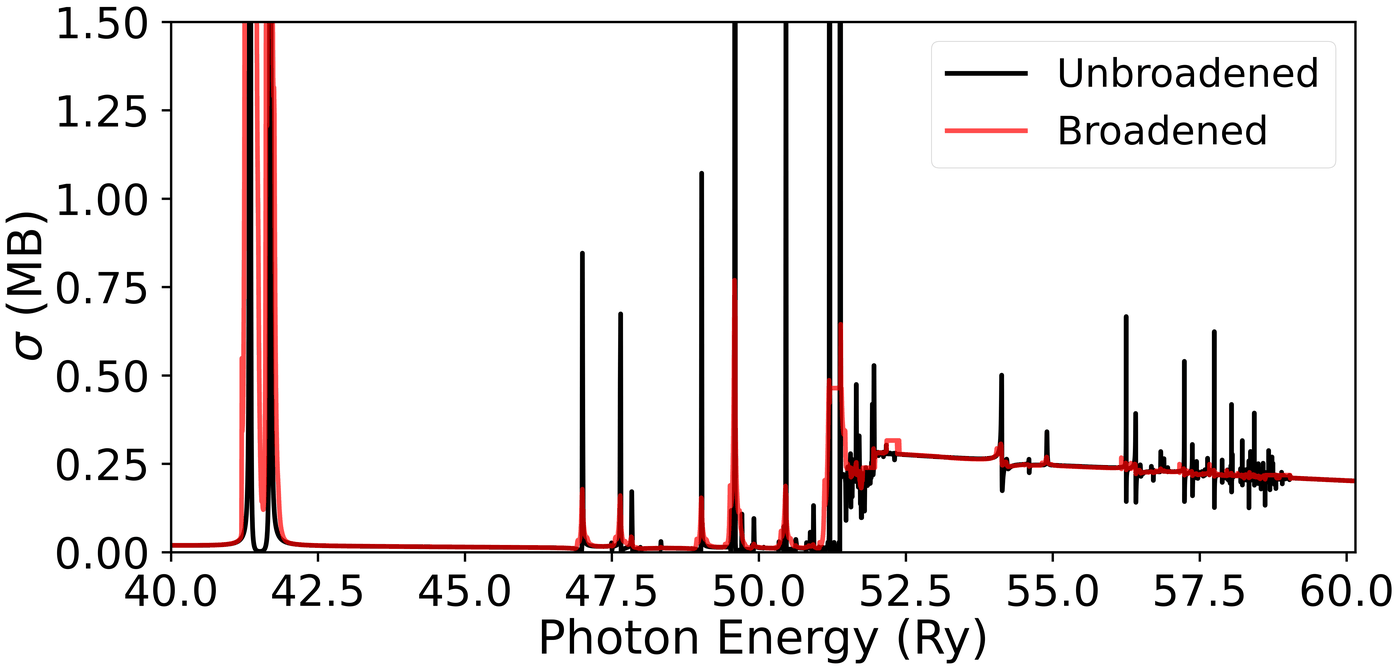}
        \caption{$N_e = 10^{18}$}
    \end{subfigure}
    \hfill
    \begin{subfigure}[b]{0.49\textwidth}
        \includegraphics[width=\textwidth]{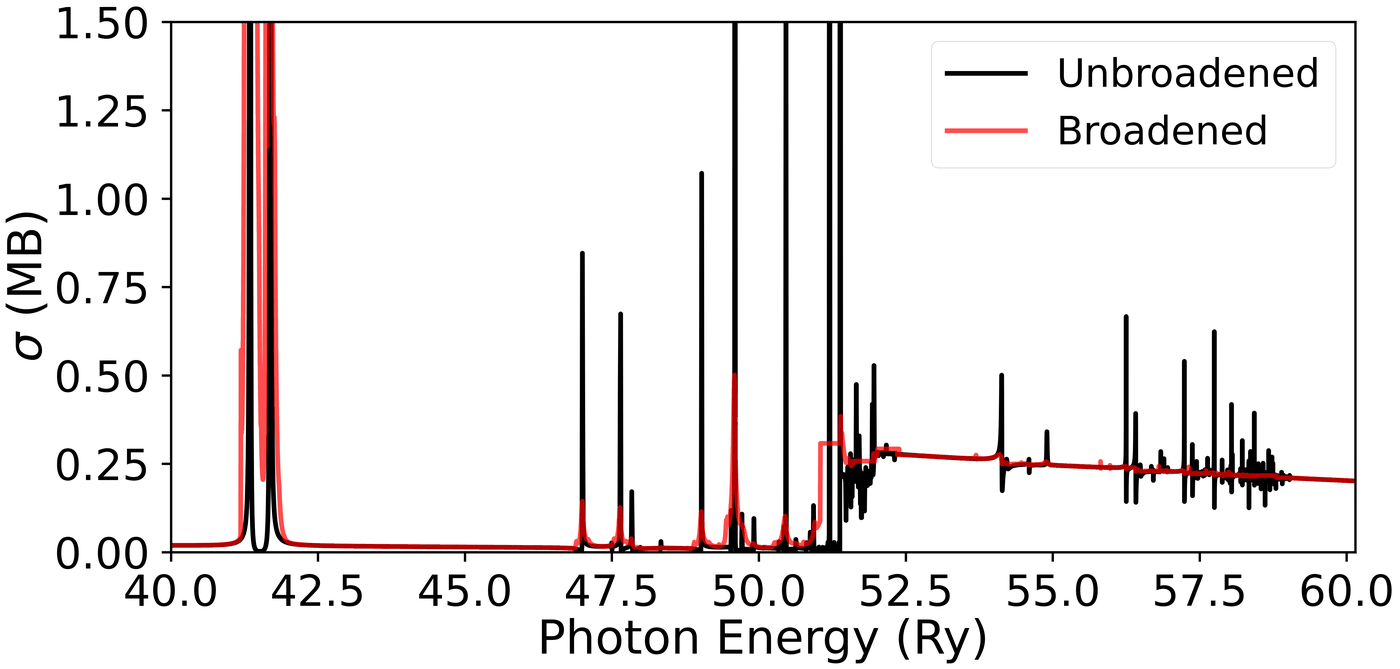}
        \caption{$N_e = 10^{19}$}
    \end{subfigure}

    \vspace{1em}

    % Row 2
    \begin{subfigure}[b]{0.49\textwidth}
        \includegraphics[width=\textwidth]{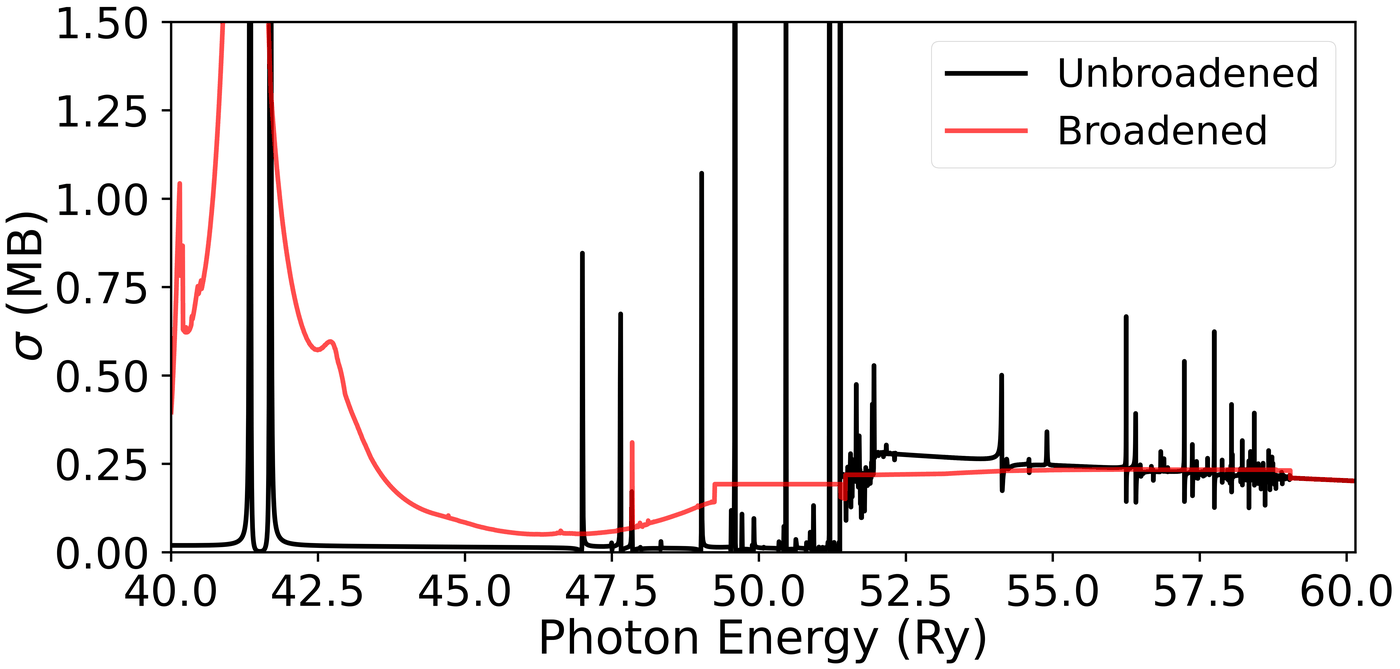}
        \caption{$N_e = 10^{22}$}
    \end{subfigure}
    \hfill
    \begin{subfigure}[b]{0.49\textwidth}
        \includegraphics[width=\textwidth]{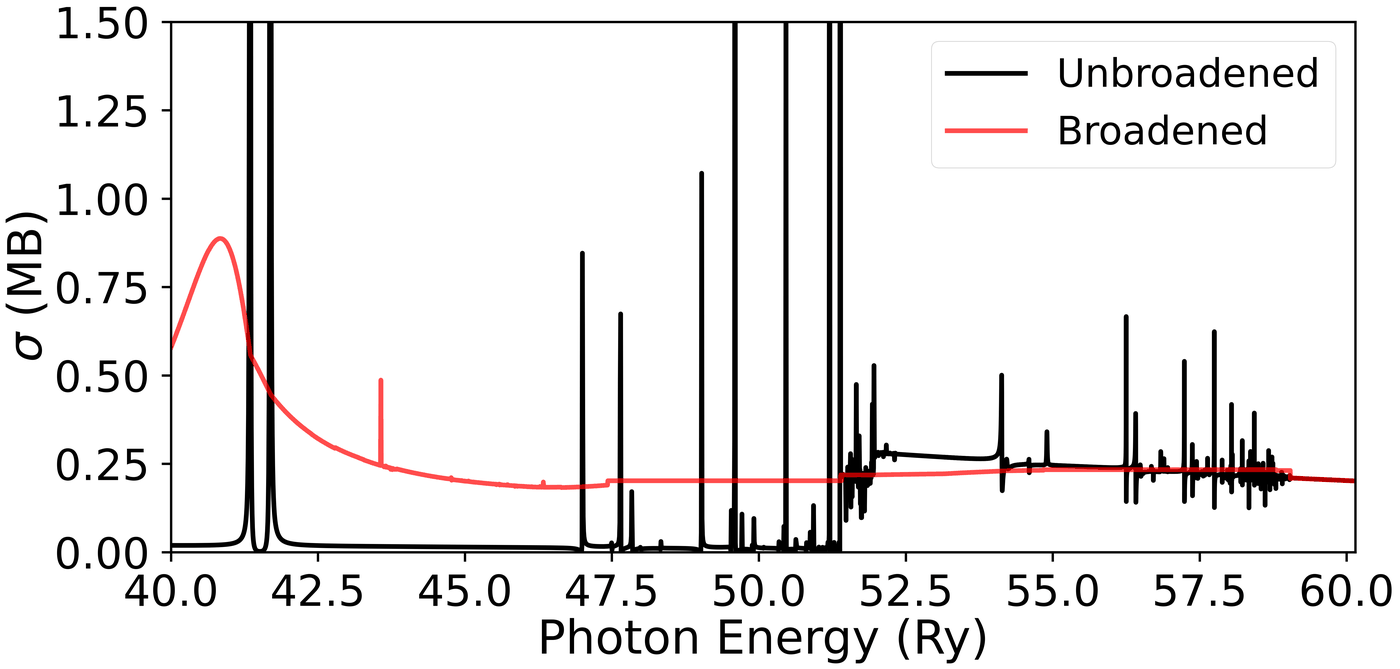}
        \caption{$N_e = 10^{23}$ BCZ}
    \end{subfigure}

    \vspace{1em}

    \caption{Linear scale: plasma broadened photoionization cross sections for 
    $\hbar\omega + \mathrm{O\,VI} \rightarrow e + \mathrm{O\,VII}$ 
    from the bound level $1s^2\,(^{1}S)\,2s\,a^{2}S_{1/2}$ at 
    $\mathbf{T =2 \times 10^{6}\,\mathrm{K}}$ for varying electron densities.
    }
    \label{fig:lin_2e6}
\end{figure}

Figures~\ref{fig:log_1e6}--\ref{fig:lin_2e6} show the plasma-broadened photoionization 
cross sections for O~\scalebox{0.8}{VI} from the ground level $1s^2\,(^{1}S)\,2s\,a^{2}S_{1/2}$ 
at temperatures $T = 1\times10^6$~K and $T = 2\times10^6$~K, across electron densities 
spanning $N_e = 10^{18}$--$10^{23}$~cm$^{-3}$. Both logarithmic and linear scales are 
presented: the log scale shows broadening across the full energy range, while the linear scale shows the actual magnitude of resonance suppression. Key density thresholds are summarized in Table~\ref{tab:broadening_limits}. The density limits correspond to: (i) the onset of resonance broadening where the stronger PEC features or wider low-$n$ Rydberg resonances are mostly unaffected, but the narrower higher-$n$ Rydberg resonances dissolve into the continua, and (ii) density where all resonances structures are flattened and thereby raise the overall continuum. It is particularly noteworthy that the density range from the onset to dissolution is roughly between $10^{20}-10^{22}/cc$. This range for plasma broadening of autoionizing resonances is different from, but consistent with, that for Fe-ions \cite{rmop3}.

{\it One of the main features of HED plasma effects is their dependence on, and delimiting values of, temperature-density ranges that are particular to each atomic ion.} At the lowest density ($N_e = 10^{18}$~cm$^{-3}$), the broadened cross section closely 
follows the unbroadened profile, with resonances remaining sharp and well-resolved. As the 
electron density increases, the autoionizing resonances progressively wash out: the sharp 
peaks and minima of the unbroadened cross section are smeared by the Lorentzian convolution, 
redistributing resonance strength into the surrounding continuum. By $N_e = 10^{22}$~cm$^{-3}$ 
the resonance structure is substantially broadened and wiped out, and at the BCZ conditions 
($N_e = 10^{23}$~cm$^{-3}$) the broadened cross section is nearly flat across the 
40--60~Ry resonance region, with individual features almost entirely dissolved into the background continuum.

The linear-scale plots in Figures~\ref{fig:lin_1e6} and~\ref{fig:lin_2e6} make the 
progression particularly clear, showing how the broadened profile transitions from 
closely tracking the resonance peaks at low density to a smooth, featureless continuum 
at BCZ conditions ($N_e = 10^{23}$~cm$^{-3}$). This systematic flattening of the 
bound-free cross section with increasing $N_e$ has direct implications for the 
monochromatic opacity, as the redistribution of resonance strength modifies the 
photon-energy dependence of opacity in the spectral region dominated by these 
autoionizing features. The progression at the two isotherms is very similar. At these conditions the total width is dominated by electron-impact broadening, a Lorentzian profile with wide wings that scales with $N_e$. Temperature enters only through the much narrower core-dominated Doppler width, which decays faster as a Gaussian. Both isotherms are retained because they correspond to the BCZ temperature range and show that the density-driven dissolution of resonances is robust across this interval.

\subsubsection{Monochromatic Opacities}

\begin{figure}[H]
 \includegraphics[width=1.0\textwidth]{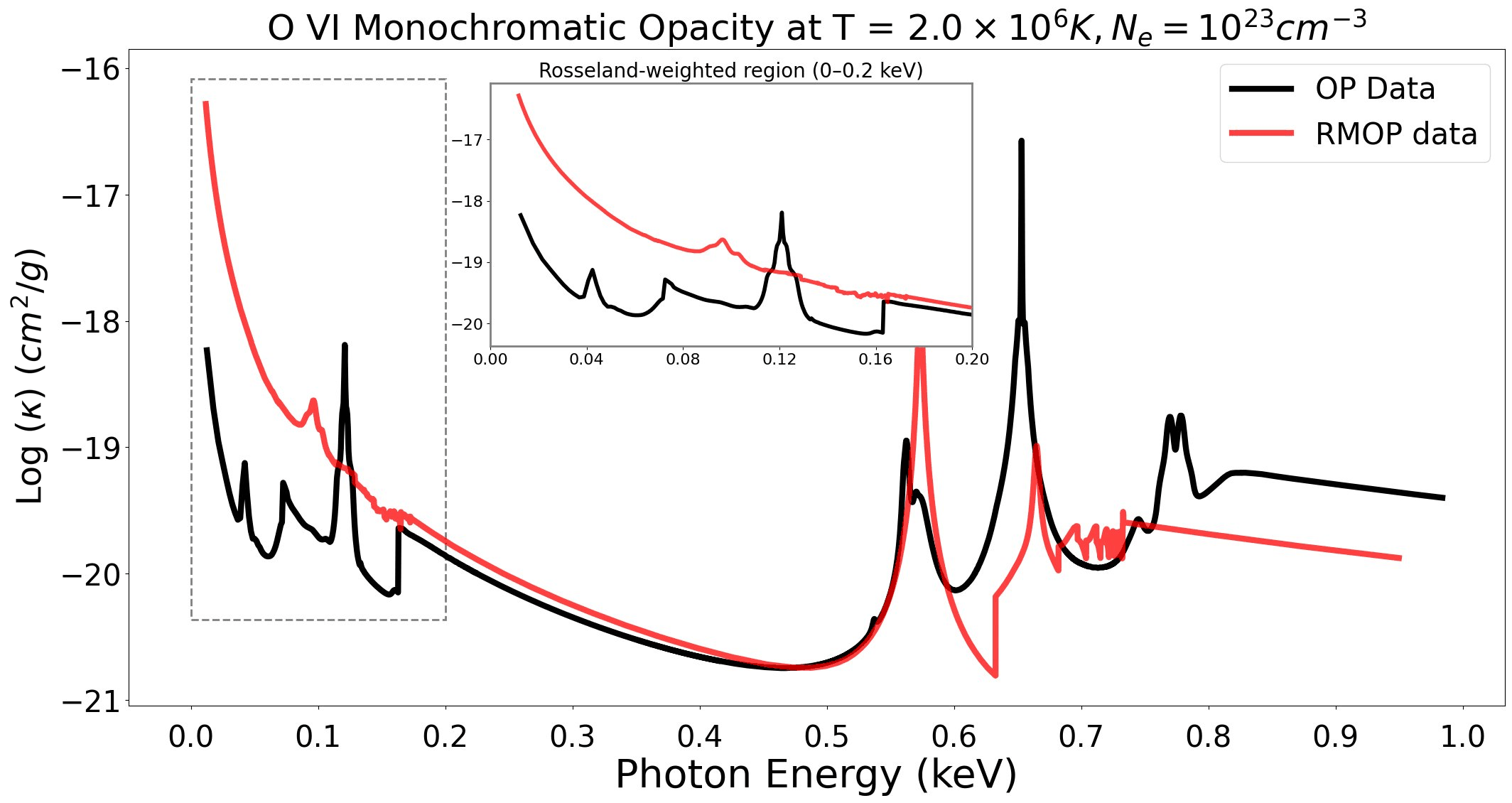}
    \caption{Logarithmic monochromatic opacities for O\,\scalebox{0.8}{VI} at BCZ conditions ($T = 2.0 \times 10^6$\,K, $N_e = 10^{23}$\,cm$^{-3}$), computed using the R-matrix method (red) and OP (black). The inset zooms on the $0$--$0.2$\,keV region, where the Planck derivative peaks at $T \sim 10^6$\,K and contributions to the Rosseland mean opacity are concentrated. RMOP yields systematically higher opacities than OP throughout this region, primarily owing to differences in photoionization cross sections.}
    \label{fig:mono-opac-6}
\end{figure}

Figure~\ref{fig:mono-opac-6} shows the monochromatic opacity $\kappa(E)$ for O\,\scalebox{0.8}{VI} at $T = 2.0 \times 10^6$~K and $N_e = 10^{23}$~cm$^{-3}$, comparing RMOP and OP results across the photon energy range $0$--$1$~keV. The inset highlights the $0$--$0.2$~keV region, where the Planck derivative peaks and contributions to the Rosseland mean opacity are concentrated at $T \sim 1$--$2 \times 10^6$~K (\eg \cite{p24}). In this region, RMOP yields systematically higher opacities than OP, with an RMOP/OP opacity ratio of mean $\sim 7$ (median $\sim 4$) across $0.02$--$0.2$~keV and local enhancements exceeding an order of magnitude at resonance features, indicating that the broad agreement at higher energies is not representative of the discrepancies relevant for stellar interior modeling. Both calculations exhibit large prominent bound-bound features near $0.55$\,keV and $0.65$\,keV; also with RMOP resonances that are sharper and shifted relative to their OP counterparts. The shifts vary across the spectrum, reflecting level-dependent channel coupling and configuration interaction effects in the BPRM treatment that are more extensive than OP \cite{snn98}. The RMOP--OP comparison shows the cumulative impact of the improved BPRM treatment rather than a one-to-one benchmark, as the two datasets differ in coupling scheme, energy coverage, and resonance treatment.

\subsection{He-like \ovii}
Unlike Li-like \ovi, for He-like \ovii the first ionization threshold lies very high in energy. Therefore, the Rydberg series of autoionization resonances converging on to \en = 2 and higher excited levels of the core ion H-like \oviii also manifest themselves relatively close to those levels, as demonstrated in the results below. The OP calculations for the helium isoelectronic sequence ions were carried out in LS coupling, and included pseudo-orbitals $\overline{1}p$ and $\overline{1}d$ for electron correlation, in addition to H-like $1s, 2s, 2p$ target orbitals \cite{Fernley_1987}. The RMOP calculations include all spectroscopic hydrogenic orbitals $1s - 3d$ \cite{snn98}.

\subsubsection{Photoionization cross sections}
\begin{figure}[H]
    \centering
    \begin{subfigure}{0.48\textwidth}
        \centering
        \includegraphics[width=\textwidth]{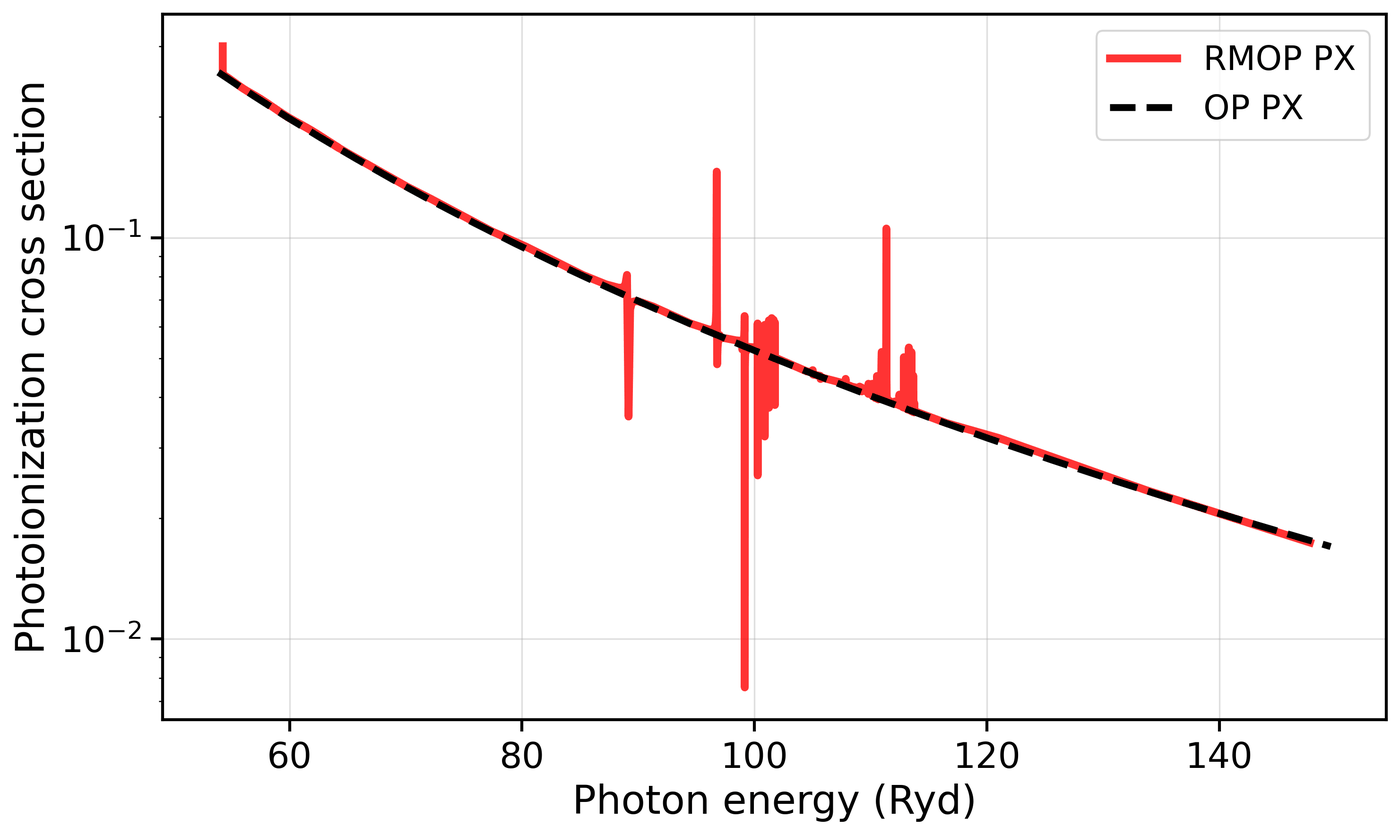}
        \caption{$1s\,(^{2}S)\,a^{1}S_0$ E = 54.3 Ry}
    \end{subfigure}
    \hfill
    \begin{subfigure}{0.48\textwidth}
        \centering
        \includegraphics[width=\textwidth]{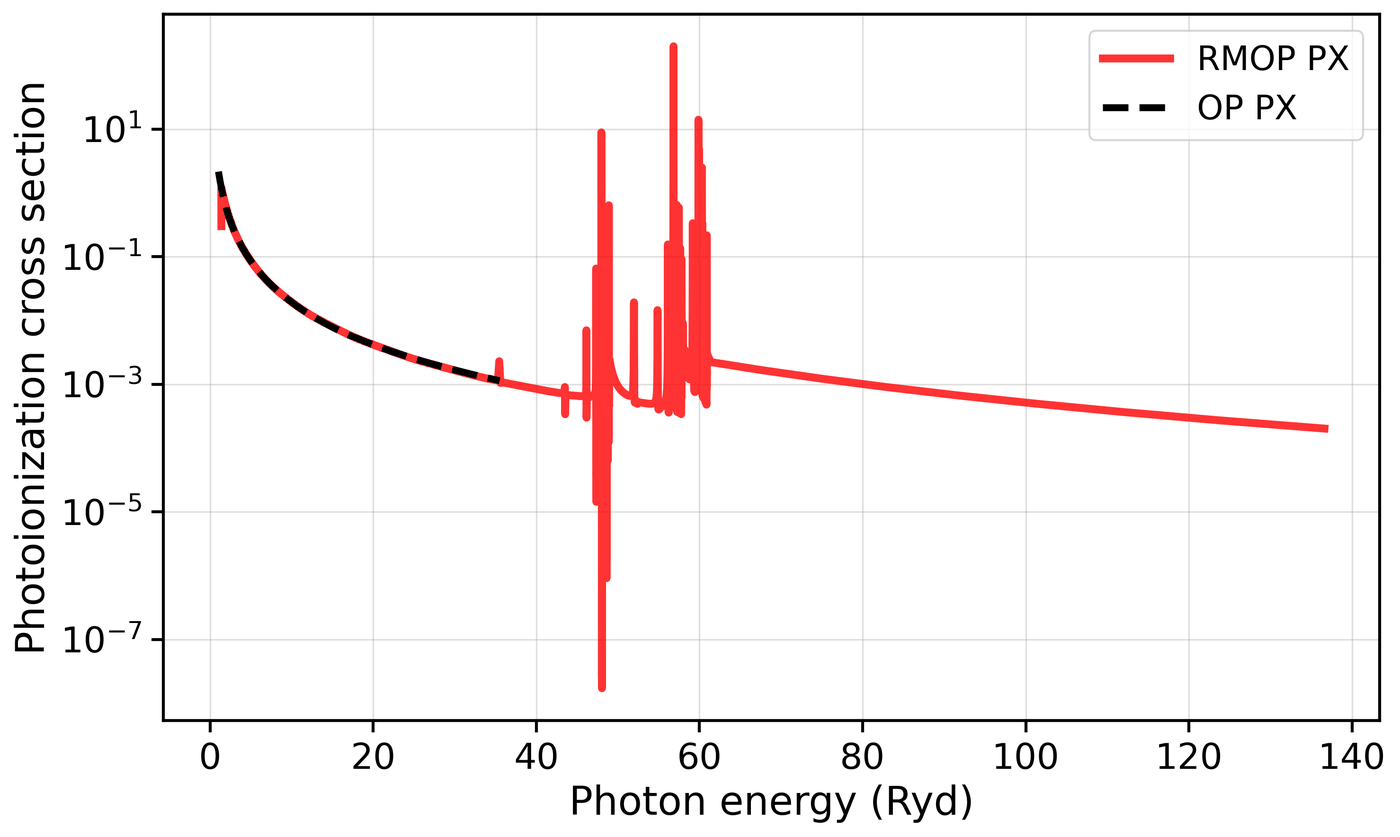}
        \caption{$1s\,(^{2}S)\,6s\,e^{3}S_1$, E = 1.38 Ry}
    \end{subfigure}

    \vspace{1em}

    \begin{subfigure}{0.48\textwidth}
        \centering
        \includegraphics[width=\textwidth]{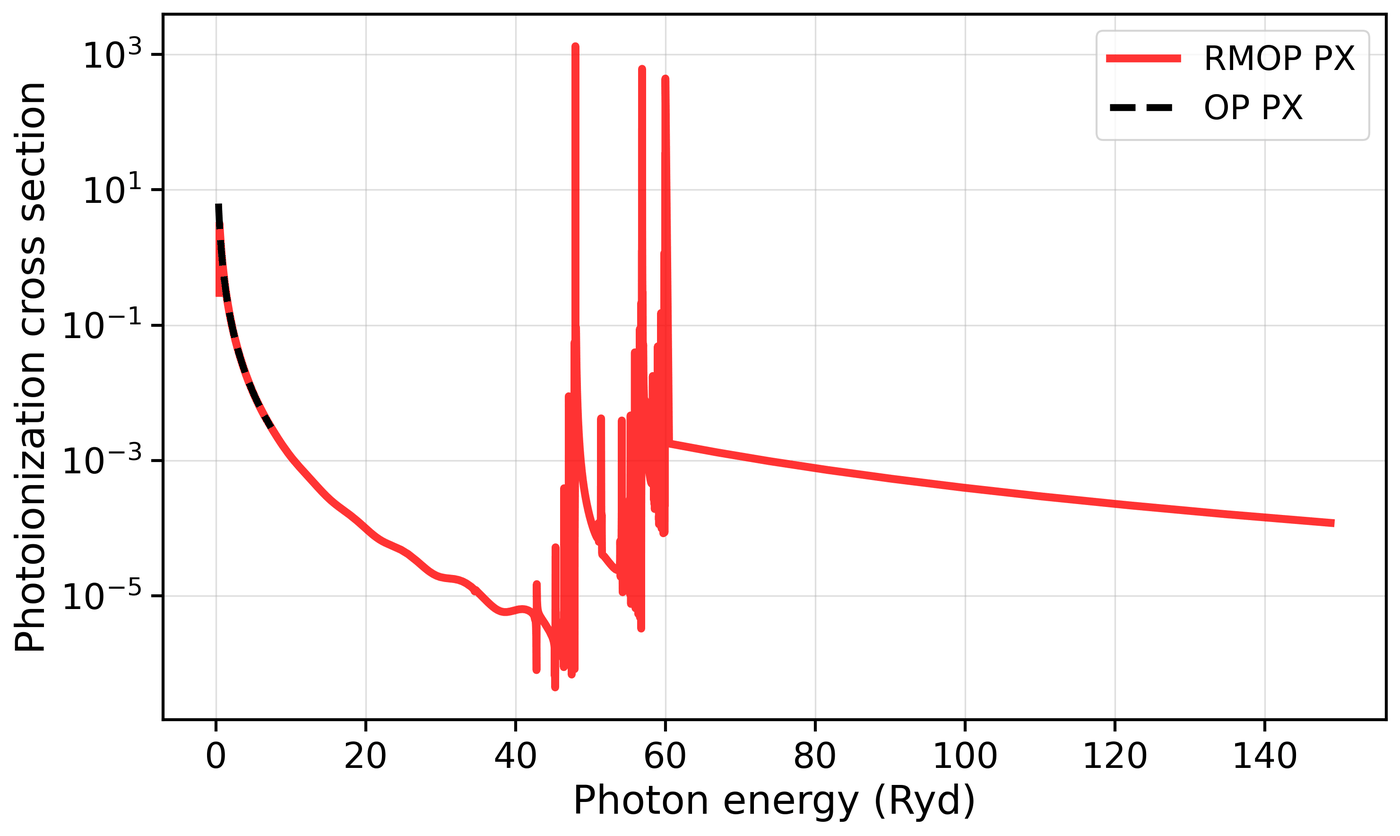}
        \caption{$1s\,(^{2}S)\,10d\,h^{3}D$, E = 0.49 Ry}
    \end{subfigure}
    \caption{Comparison of OP and RMOP photoionization cross sections for selected \ovii levels (out of 53 total). The RMOP cross sections exhibit autoionization resonance structures extending over a broad photon-energy range, while the OP data show no resonances and terminate at lower energies.}
    \label{fig:px_o7}
\end{figure}

Figure~\ref{fig:px_o7} shows BPRM photoionization cross sections for three representative 
levels of O~\scalebox{0.8}{VII}—$1s\,(^{2}S)\,a^{1}S_0$ (E = 54.3 Ry), 
$1s\,(^{2}S)\,6s\,e^{3}S_1$ (E = 1.38 Ry), and $1s\,(^{2}S)\,10d\,h^{3}D$ 
(E = 0.49 Ry)—compared with OP data. The OP database contains 53 LS levels against 
103 fine structure levels in the BPRM calculations; matching and identifying by ionization energies in the OP and RMOP datasets yields 53 common levels across which the trends shown here are representative.

Overall, the same general behaviour observed for O~\scalebox{0.8}{VI} persists for O~\scalebox{0.8}{VII}. 
Below the first excited \en = 2 core threshold both datasets are in close agreement, tracing a smooth hydrogenic background continuum. Above the core thresholds of 
O~\scalebox{0.8}{VIII}, the RMOP cross sections develop a rich autoionizing resonance 
structure with peaks spanning several orders of magnitude, while the OP data show 
no resonances and terminate at lower photon energies. For the ground state, the 
resonances are distributed over a broad photon-energy range, reflecting the 
high ionization threshold of this level. For the excited levels, the resonance 
structures concentrate in the $\sim 40$--$60$~Ry window and becomes increasingly 
dense as the level energy decreases; this is clearly seen in the progression from say, the 
$6s$ to the $10d$ level. The RMOP cross sections extend continuously to 
$\sim 700$~Ry in most cases, fully capturing the high-energy bound-free contribution absent from current OP data.

\subsubsection{Plasma broadening of photoionization cross sections}

\begin{figure}[H]
    \centering

    % Row 1
    \begin{subfigure}[b]{0.49\textwidth}
        \includegraphics[width=\textwidth]{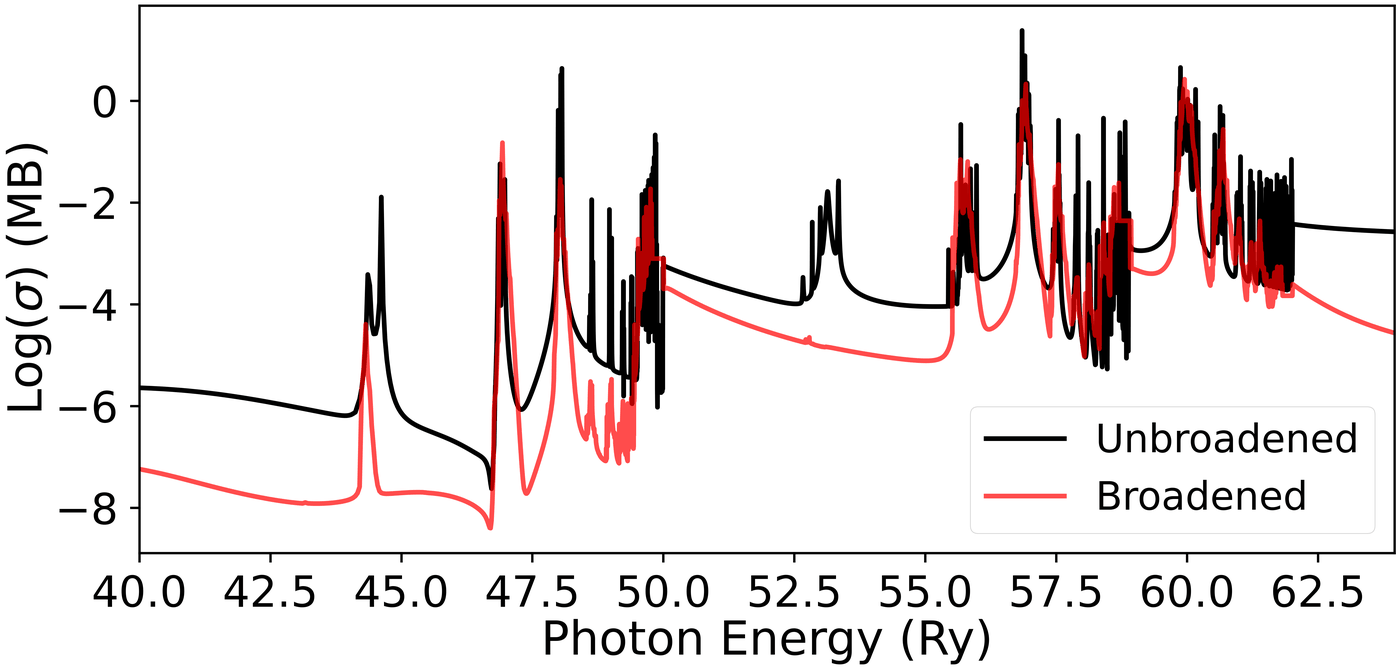}
        \caption{$N_e = 10^{18}$}
    \end{subfigure}
    \hfill
    \begin{subfigure}[b]{0.49\textwidth}
        \includegraphics[width=\textwidth]{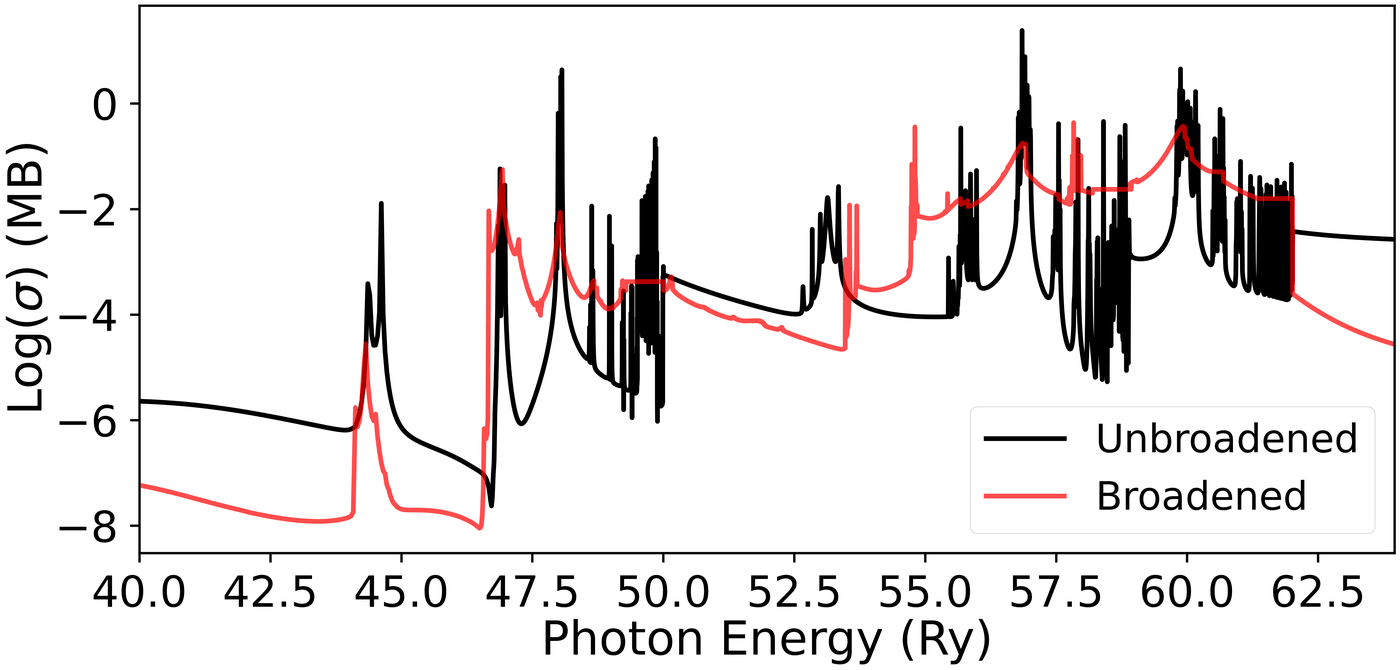}
        \caption{$N_e = 10^{20}$}
    \end{subfigure}

    \vspace{1em}

    % Row 2
    \begin{subfigure}[b]{0.49\textwidth}
        \includegraphics[width=\textwidth]{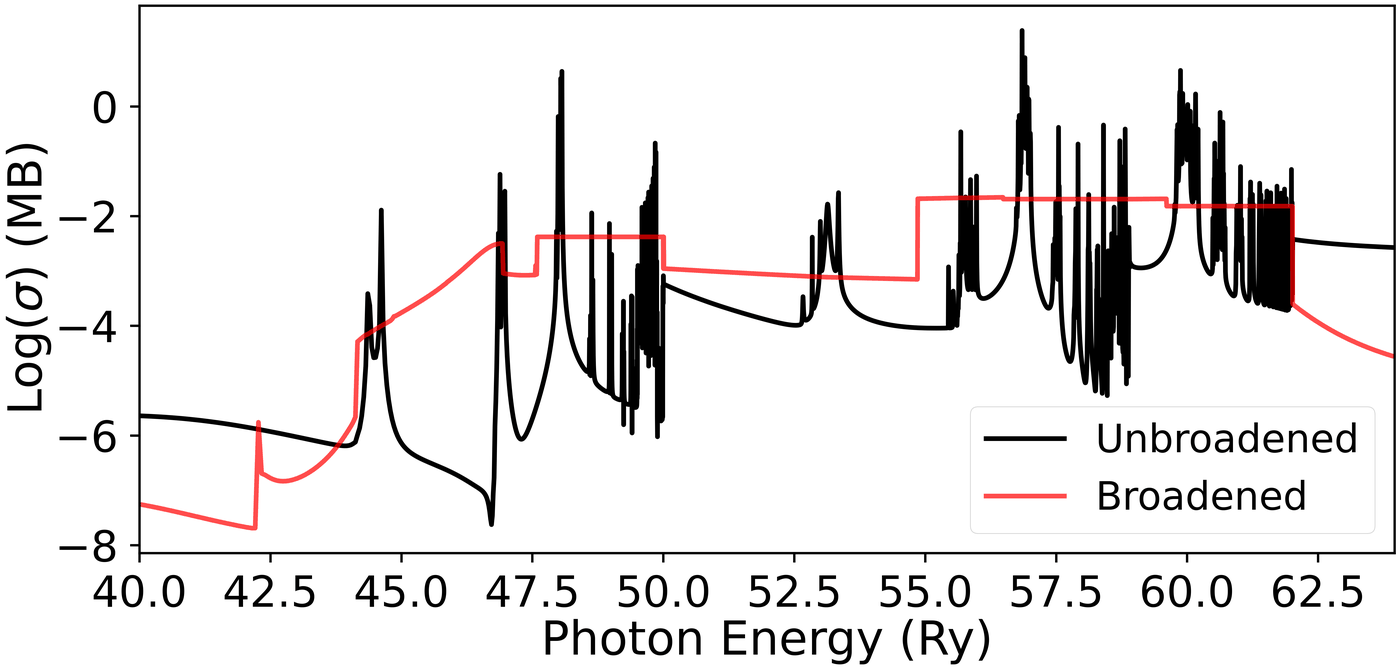}
        \caption{$N_e = 10^{22}$}
    \end{subfigure}
    \hfill
    \begin{subfigure}[b]{0.49\textwidth}
        \includegraphics[width=\textwidth]{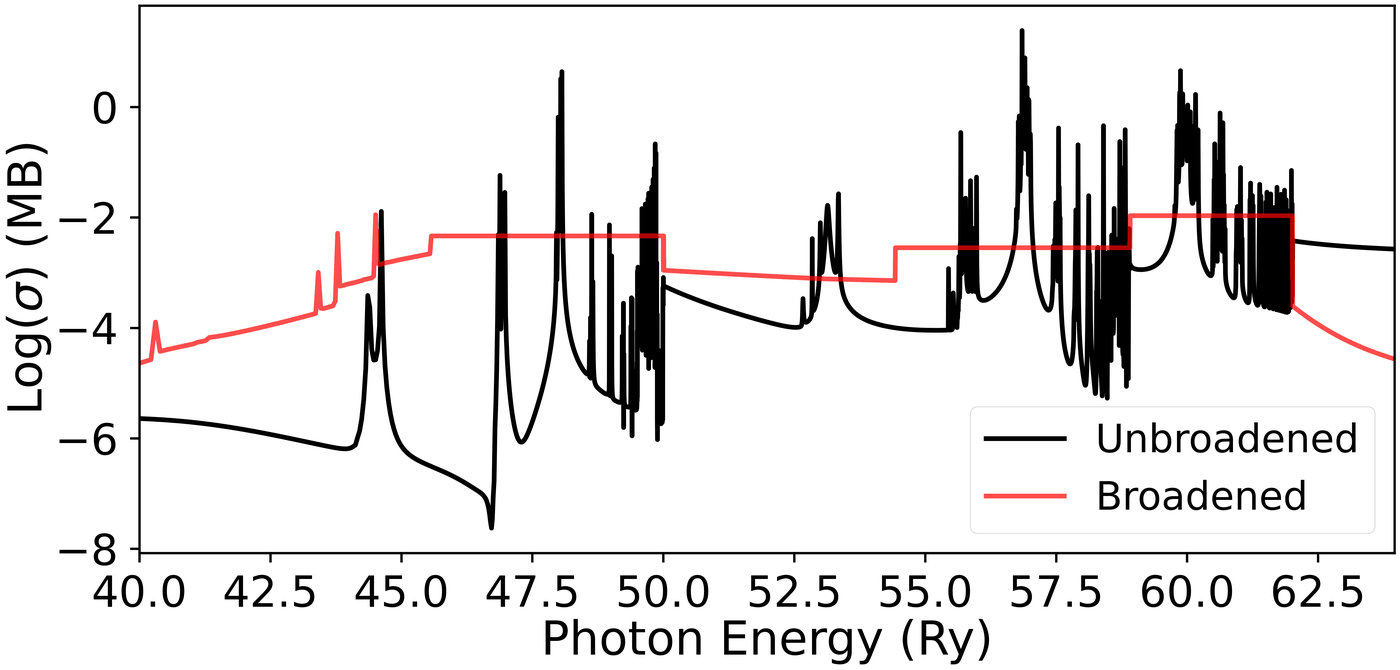}
        \caption{$N_e = 10^{23}$}
    \end{subfigure}

    \vspace{1em}

    \caption{Logarithmic plasma broadened photoionization cross sections for 
    $\hbar\omega + \mathrm{O\,VII} \rightarrow e + \mathrm{O\,VIII}$ 
    from the bound level $1s\,4f\ ({}^{1}F^{o}_{3})$ at 
    $\mathbf{T = 1 \times 10^{6}\,\mathrm{K}}$ for varying electron densities.
    }
    \label{fig:o7_log_1e6}
\end{figure}

\begin{figure}[H]
    \centering
    % Row 1
    \begin{subfigure}[b]{0.49\textwidth}
        \includegraphics[width=\textwidth]{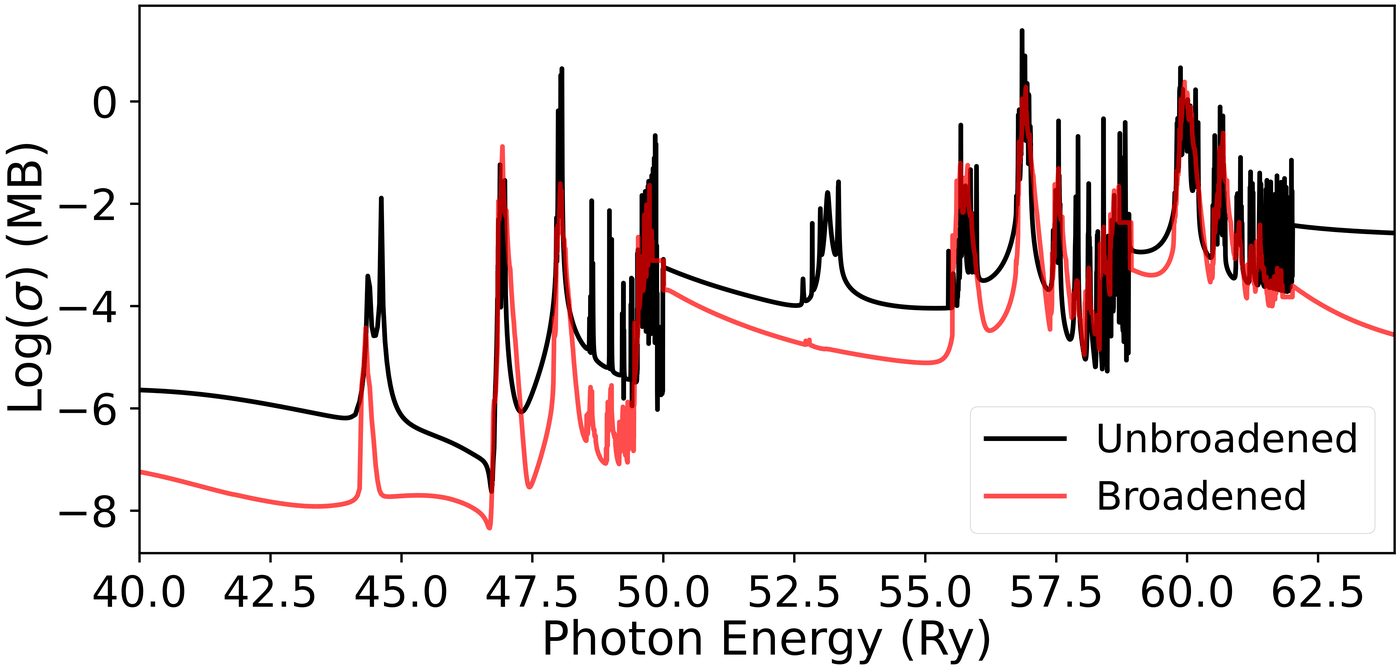}
        \caption{$N_e = 10^{18}$}
    \end{subfigure}
    \hfill
    \begin{subfigure}[b]{0.49\textwidth}
        \includegraphics[width=\textwidth]{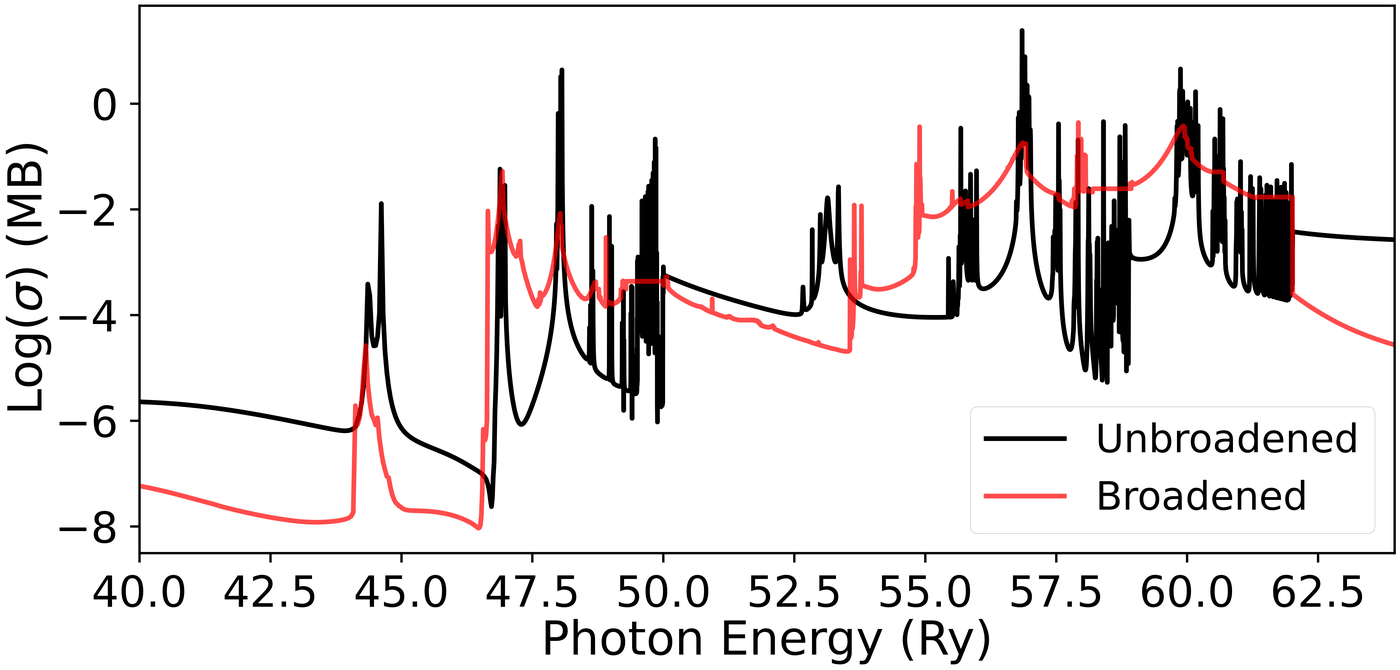}
        \caption{$N_e = 10^{20}$}
    \end{subfigure}

    \vspace{1em}

    % Row 2
    \begin{subfigure}[b]{0.49\textwidth}
        \includegraphics[width=\textwidth]{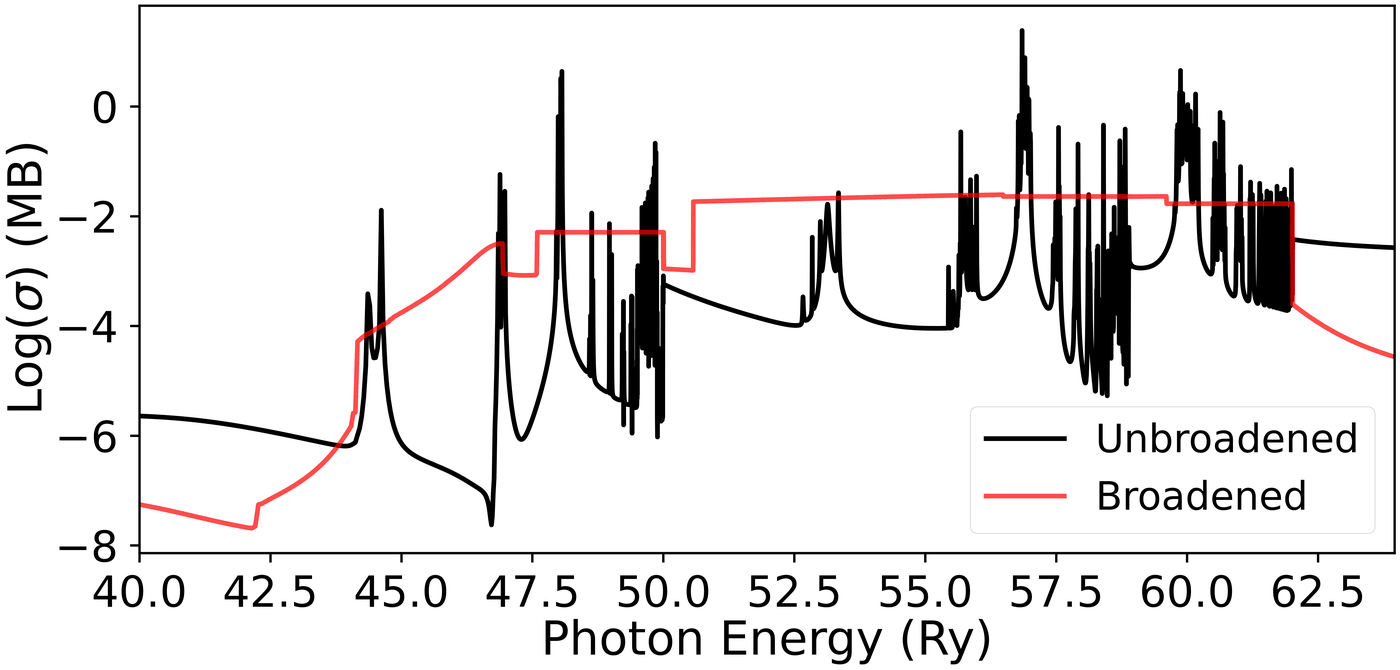}
        \caption{$N_e = 10^{22}$}
    \end{subfigure}
    \hfill
    \begin{subfigure}[b]{0.49\textwidth}
        \includegraphics[width=\textwidth]{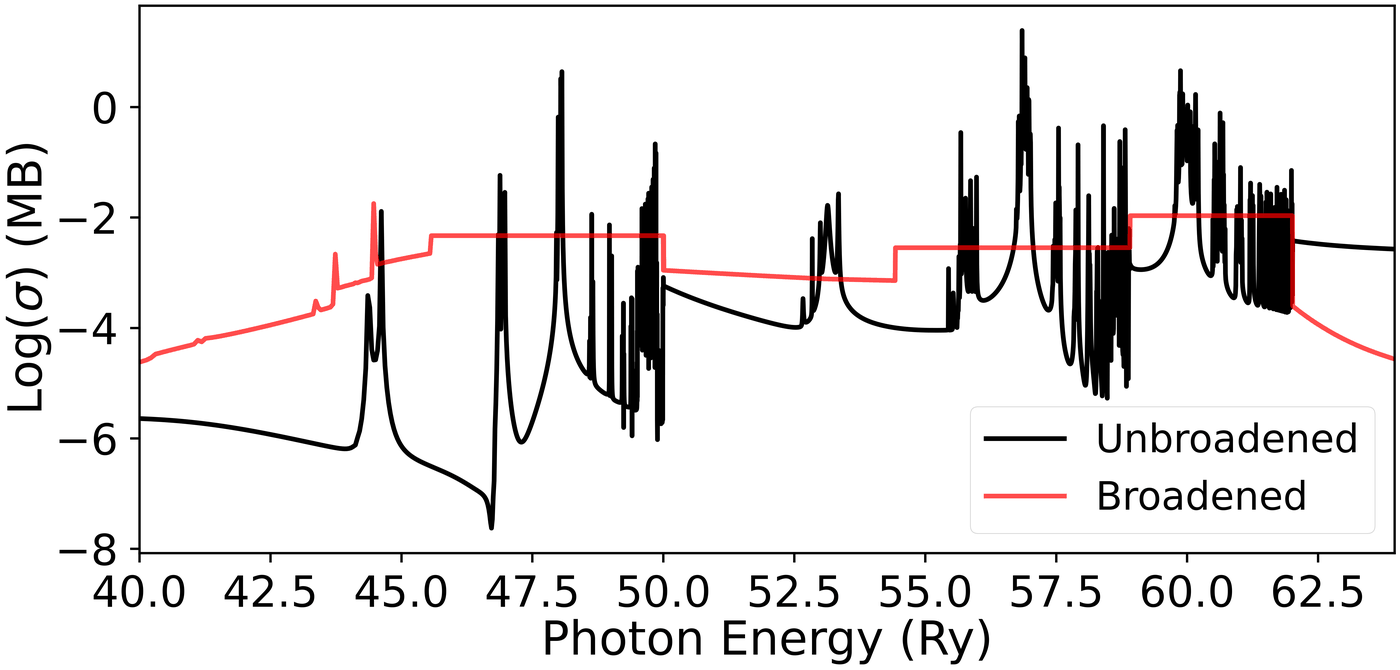}
        \caption{$N_e =10^{23}$ (BCZ)}
    \end{subfigure}

    \vspace{1em}

    \caption{ Logarithmic plasma broadened photoionization cross sections for 
    $\hbar\omega + \mathrm{O\,VII} \rightarrow e + \mathrm{O\,VIII}$ 
    from the bound level $1s\,4f\ ({}^{1}F^{o}_{3})$ at 
    $\mathbf{T = 2.0 \times 10^{6}\,\mathrm{K}}$  for varying electron densities.}
    \label{fig:o7_log_2e6}
\end{figure}

\begin{figure}[H]
    
   % Row 2
\begin{subfigure}[b]{0.49\textwidth}
    \includegraphics[width=\textwidth]{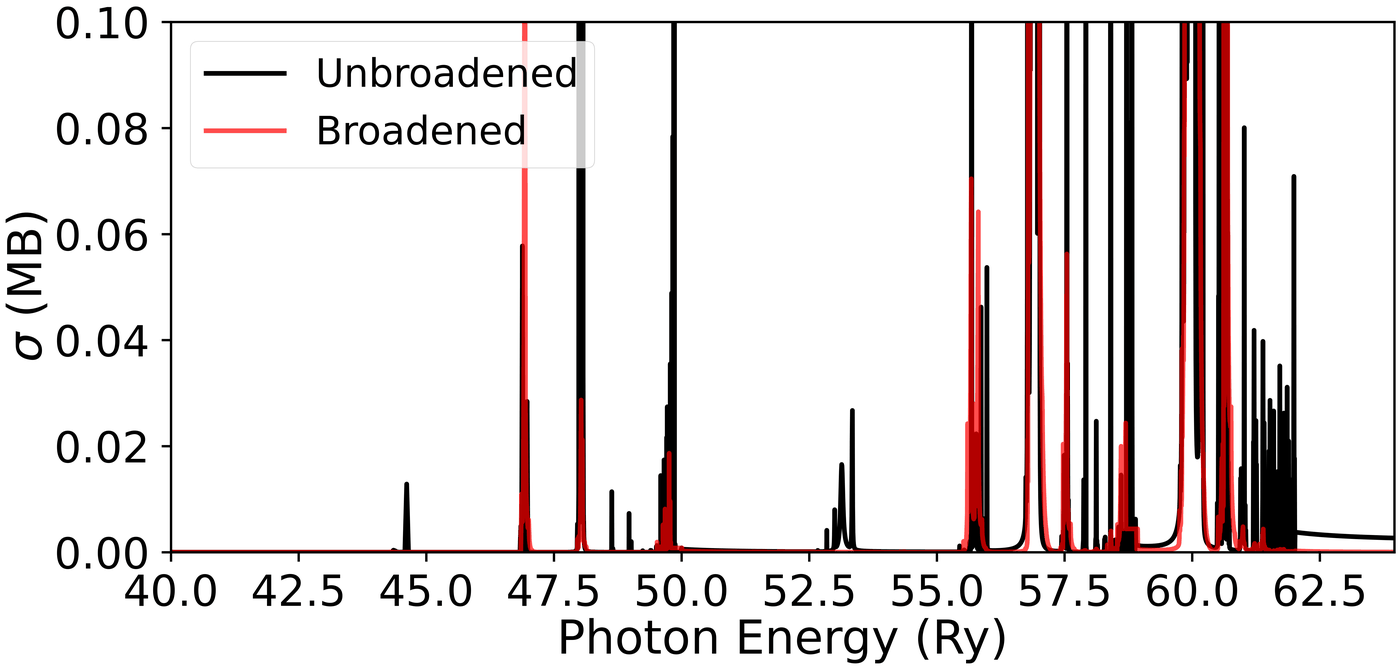}
    \caption{$N_e = 10^{18}$}
\end{subfigure}
\hfill
\begin{subfigure}[b]{0.49\textwidth}
    \includegraphics[width=\textwidth]{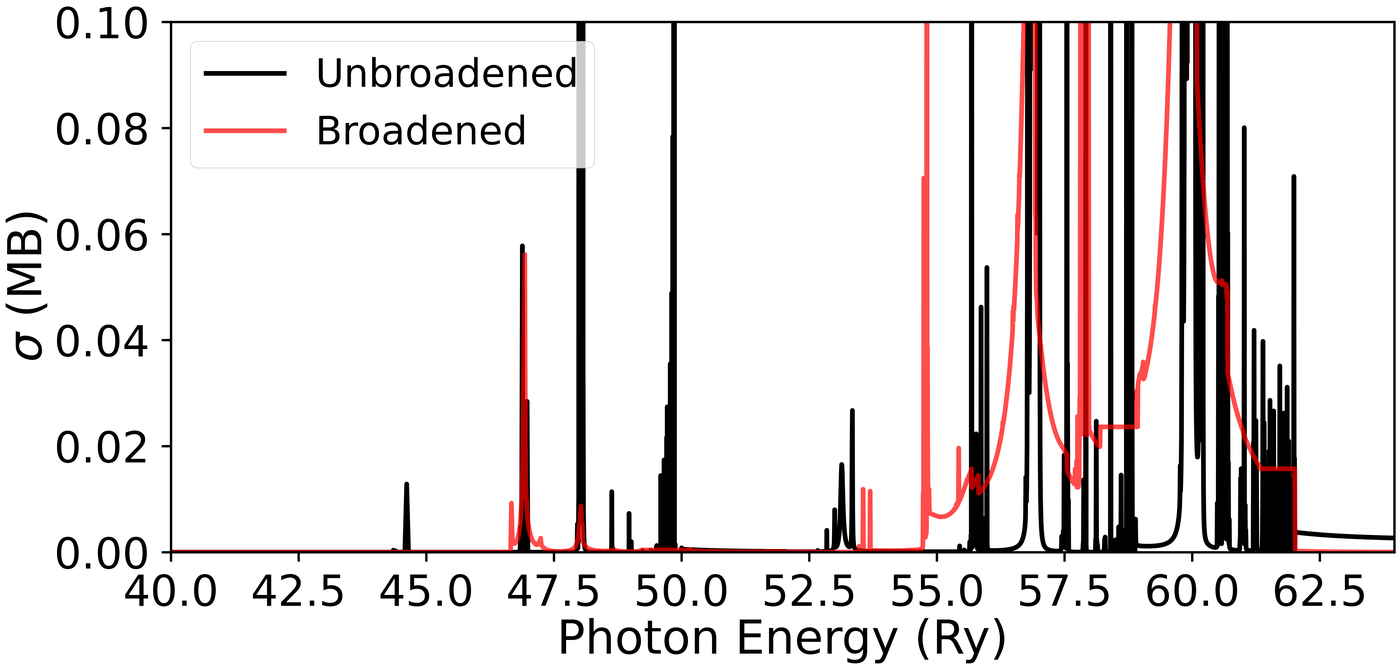}
    \caption{$N_e = 10^{20}$}
\end{subfigure}

\vspace{0.5em}

% Row 3
\begin{subfigure}[b]{0.49\textwidth}
    \includegraphics[width=\textwidth]{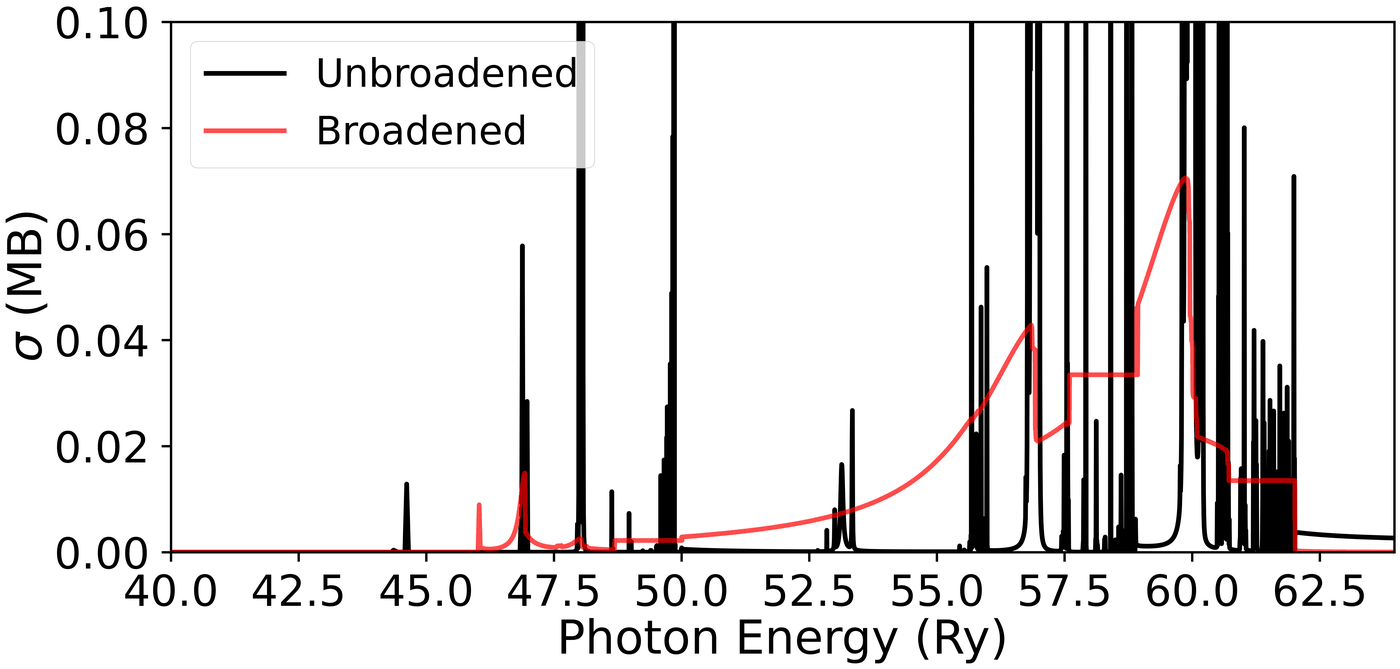}
    \caption{$N_e = 10^{21}$}
\end{subfigure}
\hfill
\begin{subfigure}[b]{0.49\textwidth}
    \includegraphics[width=\textwidth]{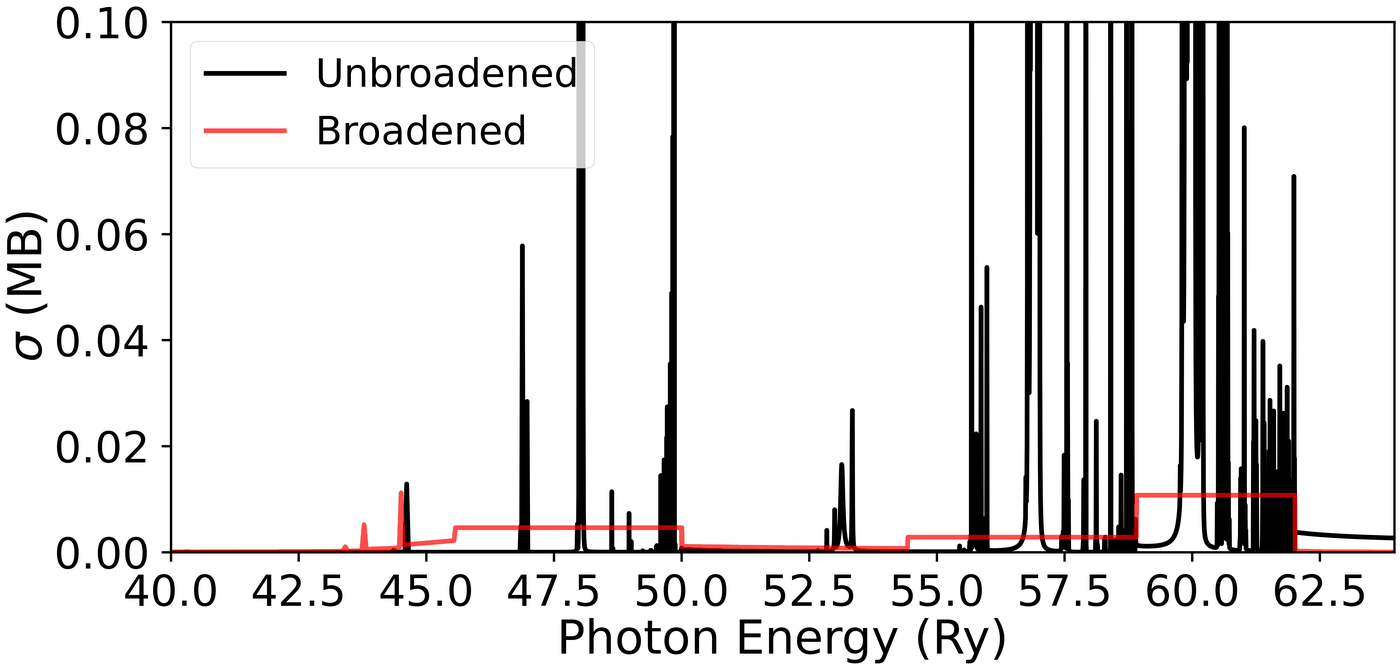}
    \caption{$N_{e} = 10^{23}$}
\end{subfigure}
\caption{Plasma broadened photoionization cross sections for 
$\hbar\omega + \mathrm{O\,VII} \rightarrow e + \mathrm{O\,VIII}$ 
from the bound level $1s\,4f\ ({}^{1}F^{o}_{3})$ at 
$\mathbf{T = 1 \times 10^{6}\,\mathrm{K}}$ for varying electron densities.
}
\label{fig:o7_lin_1e6}
\end{figure}

\begin{figure}[H]
    \centering

    % Row 1
    \begin{subfigure}[b]{0.49\textwidth}
        \includegraphics[width=\textwidth]{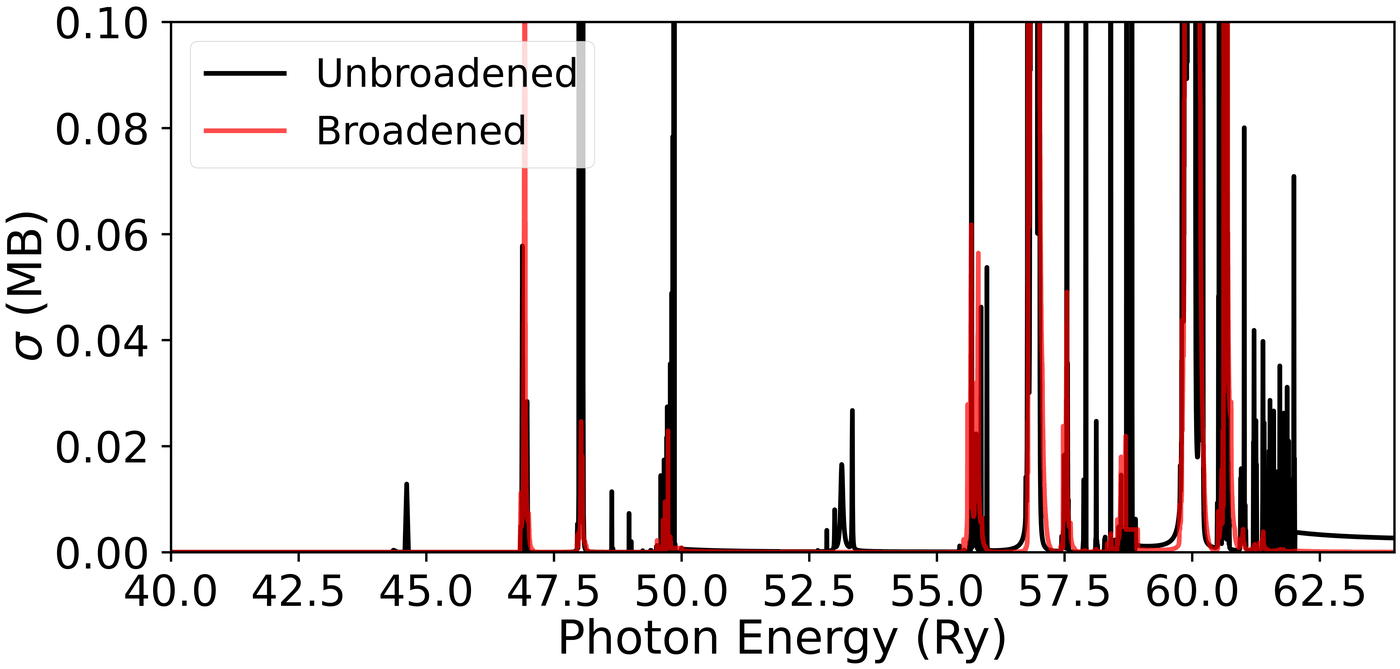}
        \caption{$N_e = 10^{18}$}
    \end{subfigure}
    \hfill
    \begin{subfigure}[b]{0.49\textwidth}
        \includegraphics[width=\textwidth]{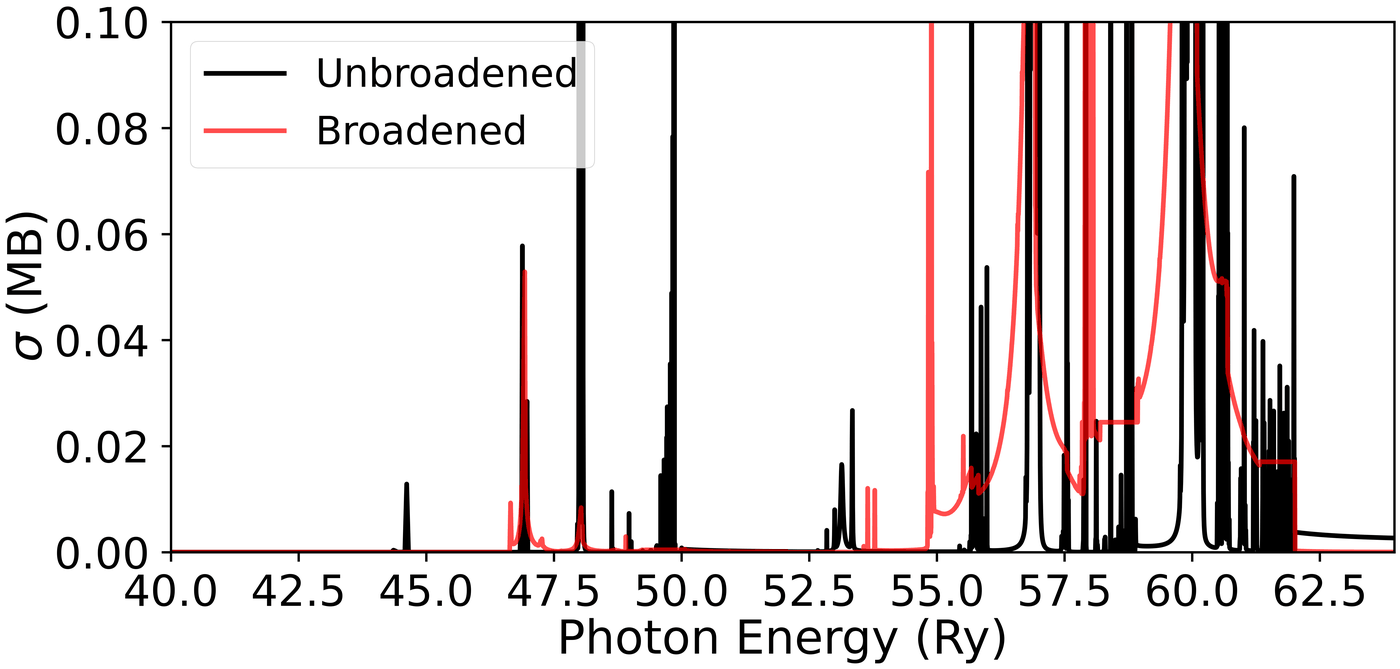}
        \caption{$N_e = 10^{20}$}
    \end{subfigure}

    \vspace{1em}

    % Row 2
    \begin{subfigure}[b]{0.49\textwidth}
        \includegraphics[width=\textwidth]{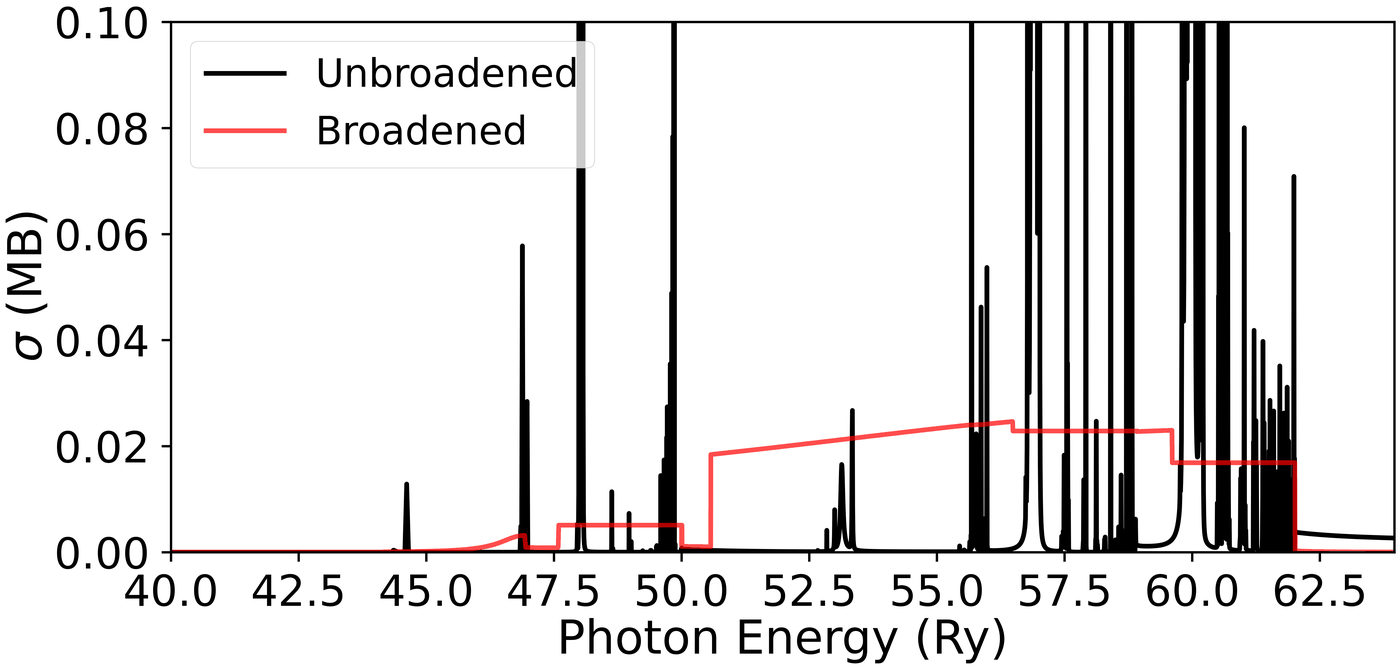}
        \caption{$N_e = 10^{22}$}
    \end{subfigure}
    \hfill
    \begin{subfigure}[b]{0.49\textwidth}
        \includegraphics[width=\textwidth]{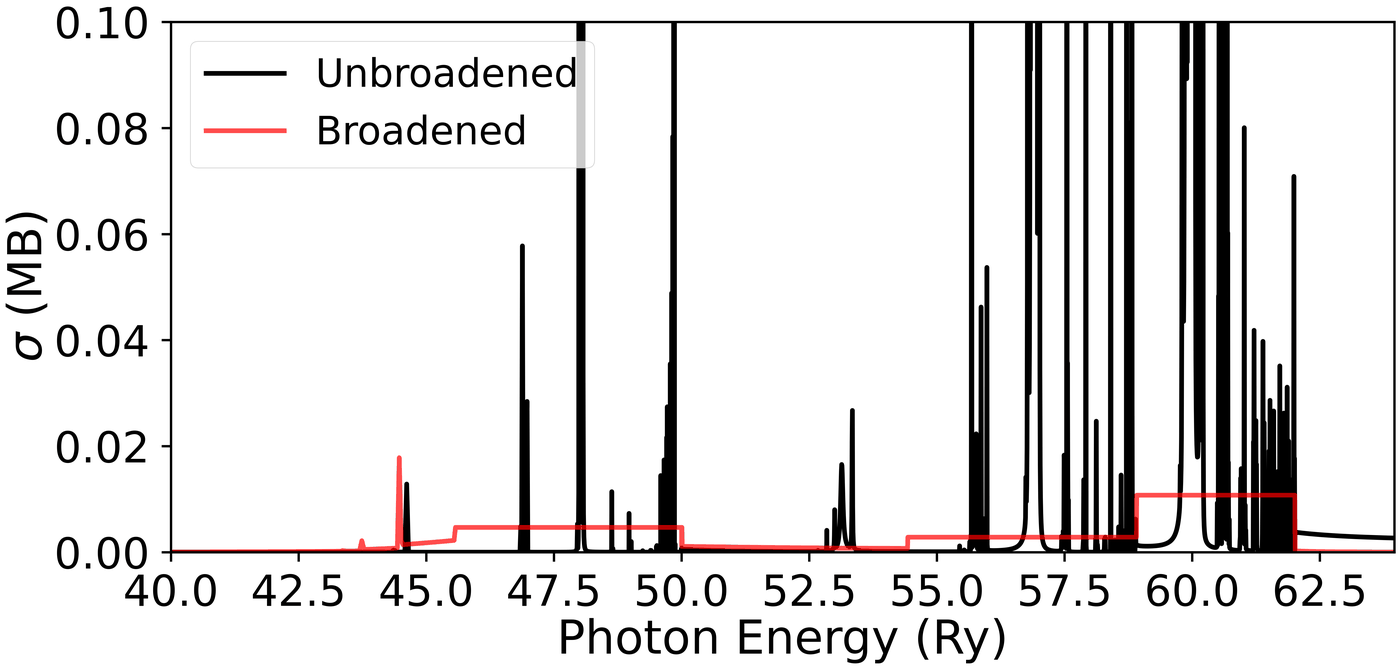}
        \caption{$N_e = 10^{23}$ (BCZ)}
    \end{subfigure}

    \vspace{1em}

    \caption{Plasma broadened photoionization cross sections for 
$\hbar\omega + \mathrm{O\,VII} \rightarrow e + \mathrm{O\,VIII}$ 
from the bound level $1s\,4f\ ({}^{1}F^{o}_{3})$ at 
$\mathbf{T = 2 \times 10^{6}\,\mathrm{K}}$ for varying electron densities. 
}
\label{fig:o7_lin_2e6}
  
\end{figure}

For the broadening study we use the $1s\,4f\,(^{1}F^{\circ}_3)$ level, whose dense autoionizing resonances in the 40--63\,Ry window clearly show the density-dependent progression. This level is part of our full set of 103 RMOP levels. Its unbroadened cross section is shown as the black curve in each panel of Figs.~\ref{fig:o7_log_1e6}--\ref{fig:o7_lin_2e6}. These figures show the plasma-broadened photoionization cross sections at $T = 1\times10^6$~K and $T = 2\times10^6$~K, across electron densities $N_e = 10^{18}$--$10^{23}$~cm$^{-3}$, including BCZ conditions ($N_e = 10^{23}$~cm$^{-3}$).

The same general broadening progression observed for O~\scalebox{0.8}{VI} is seen here: 
at low densities ($N_e = 10^{18}$~cm$^{-3}$) the broadened cross section closely 
follows the unbroadened profile, with the dense autoionizing resonance structure 
remaining well-resolved. As $N_e$ increases, the Lorentzian convolution 
progressively smears the resonances, redistributing their strength into the 
surrounding continuum until the profile flattens at BCZ conditions. The linear-scale 
plots make this progression particularly clear, showing the gradual suppression of 
individual resonance features with increasing density. In the logarithmic plots, the broadened cross section appears below the unbroadened profile in certain energy regions. This is not a numerical artifact but a consequence of the Lorentzian convolution redistributing resonance strength from sharp peaks into broader wings, which lowers peak values while raising the surrounding continuum. The linear-scale plots (Figs.~\ref{fig:o7_lin_1e6}--\ref{fig:o7_lin_2e6}) show this redistribution more faithfully, since the log scale visually exaggerates the drop at peak centers.

A notable feature of the O~\scalebox{0.8}{VII} broadening is the differential response of the two main resonance clusters. The cluster in the $\sim$47--50\,Ry region flattens at lower densities than the cluster in the $\sim$57--60\,Ry region, which retains more of its resonance structure and persists to higher $N_e$ before being fully smeared. We attribute this tentatively to the different widths and spacings of the autoionizing resonances converging to the respective O~\scalebox{0.8}{VIII} core thresholds in each energy window, with the more closely spaced resonances in the lower-energy cluster being more susceptible to overlap and dissolution at moderate densities.

Table~\ref{tab:broadening_limits} summarizes the approximate density limits at which broadening first becomes apparent (resonance peaks visibly reduced) and at which the resonance structure is largely dissolved into the continuum, for both ions and their main resonance clusters.

% ========================
% Figure 1: O7 Level 1s4f — 240
% ========================

% ========================
% Figure 2: O7 Level 1s4f — 252
% ========================

\begin{table}[H]
\centering
\caption{Approximate density limits for the onset and dissolution of resonance broadening at $T \sim 1$--$2 \times 10^6$~K. ``Onset'' denotes the density at which broadening becomes noticeably apparent in cross sections; ``dissolved'' denotes the density at which resonance structures are largely flattened into the continuum.}
\label{tab:broadening_limits}
\renewcommand{\arraystretch}{1.8}
\setlength{\tabcolsep}{18pt}
\begin{tabular}{l c c}
\hline
Ion / cluster & Onset $N_e$ (cm$^{-3}$) & Dissolved $N_e$ (cm$^{-3}$) \\
\hline
O\,\scalebox{0.8}{VI} (40--60 Ry resonances) & $\sim 10^{20}$ & $\sim 10^{23}$ \\
O\,\scalebox{0.8}{VII} (47--50 Ry cluster) & $\sim 10^{20}$ & $\sim 10^{22}$ \\
O\,\scalebox{0.8}{VII} (57--60 Ry cluster) & $\sim 10^{21}$ & $\sim 10^{23}$ \\
\hline
\end{tabular}
\end{table}

\subsubsection{Monochromatic Opacities}
\begin{figure}[H]
 \includegraphics[width=1.0\textwidth]{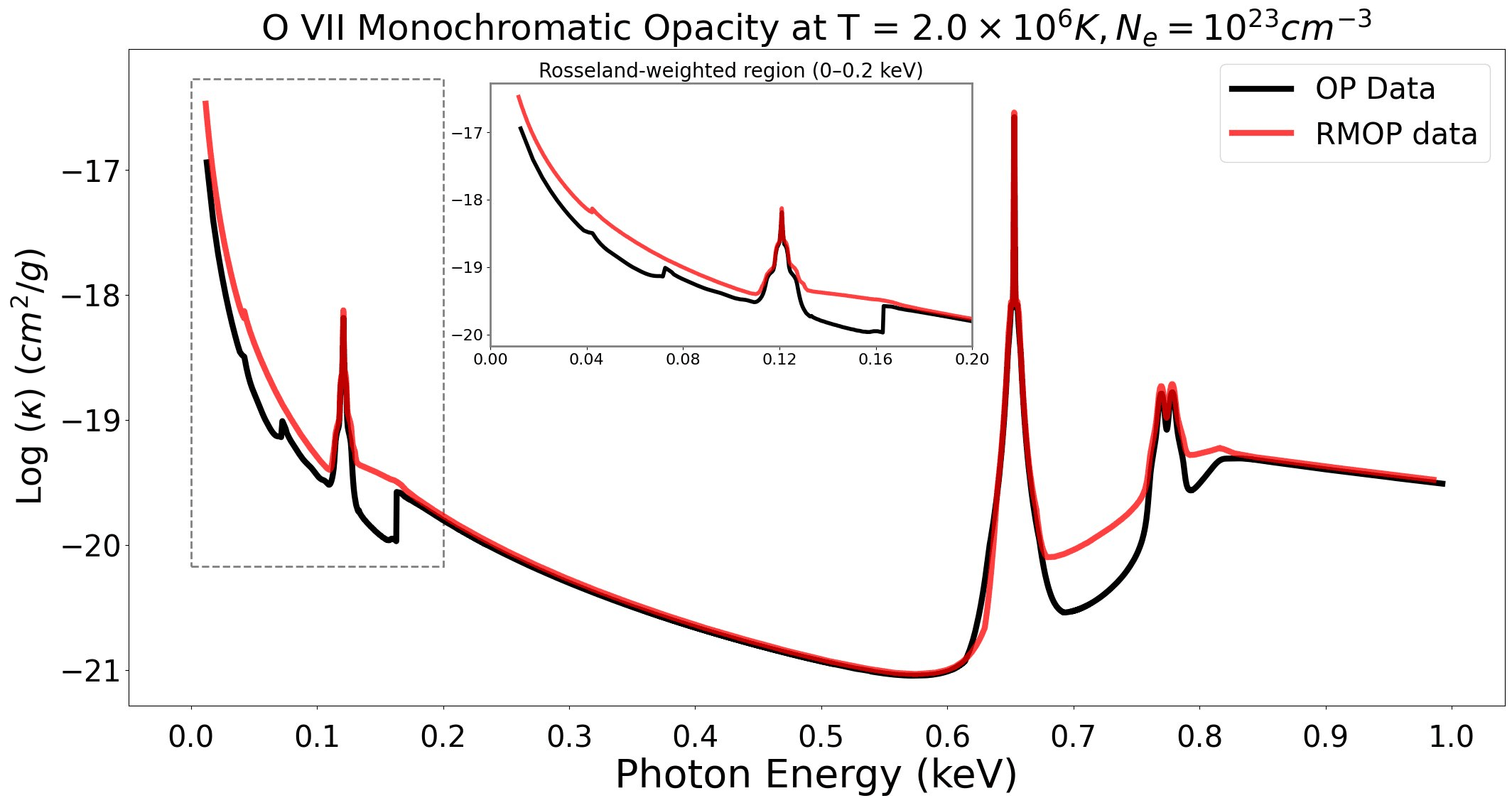}
    \caption{Monochromatic opacity spectrum of \ovii up to 1000 eV including both bound-bound and bound-free opacities. The inset region highlights the lower energy region that dominates the Rosseland Mean under BCZ conditions, and shows significant enhancement in background opacity in RMOP over OP results primarily owing to differences in photoionization cross sections.}
    \label{fig:mono-opac-7}
\end{figure}

Figure~\ref{fig:mono-opac-7} shows the monochromatic opacity $\kappa(E)$ for O\,\scalebox{0.8}{VII} at $T = 2.0 \times 10^6$~K and $N_e = 10^{23}$~cm$^{-3}$. As with O\,\scalebox{0.8}{VI}, the inset highlights the Planck derivative weighted $0$--$0.2$~keV region, where RMOP yields broadly higher opacities than OP, with an RMOP/OP opacity ratio of mean $\sim 1.4$ (median $\sim 1.2$) across $0.02$--$0.2$~keV and local enhancements up to $\sim 3 \times$ at resonance features, also due to plasma broadening effects which dissolve resonance structure and raise the background opacity. At higher energies, both calculations reproduce the same general continuum shape and prominent features near $0.65$\,keV and $0.78$\,keV, that also include the bound-bound contributions, with RMOP peaks significantly higher and sharper than their OP counterparts (the BPRM calculations include considerably more correlation effects than OP \cite{snn98}). The peak differences are less pronounced than for O\,\scalebox{0.8}{VI}, though both ions show systematic differences in the low-energy region relevant to the Planck derivative weighted Rosseland Mean Opacity that primarily governs radiation transport in HED plasmas (\eg \cite{p24,pn25}). The larger enhancement for O\,\scalebox{0.8}{VI} (median $\sim 4$) versus O\,\scalebox{0.8}{VII} (median $\sim 1.2$) reflects the stronger role of channel coupling and configuration interaction in the more complex Li-like ion.

\section{Conclusion}

We have presented R-matrix calculations for opacities (RMOP) for O\,\scalebox{0.8}{VI} and O\,\scalebox{0.8}{VII} ions, prevalent at high temperatures and densities in the solar interior close to the radiation-convection zone boundary. As an exemplar of HED sources in general, these include plasma-broadened photoionization cross sections and monochromatic opacities across wide photon energy ranges. The cross sections incorporate detailed resonance structures with plasma broadening incorporated via electron-impact (collisional), Stark, Doppler, and free-free mechanisms. We also delimit the electron density range along two isotherms for both \ovi and \ovii. Comparison with OP results shows larger discrepancies for O\,\scalebox{0.8}{VI}, where RMOP yields systematically higher opacities at low energies and sharper, shifted resonance features, whereas O\,\scalebox{0.8}{VII} shows generally good agreement between the two frameworks, with smaller differences in peak sharpness and in the low-energy continuum.

Recently, total oxygen opacity have been reported at the Sandia Z ICF facility under HED conditions \cite{Bailey-oxy25}. However, individual contributions of oxygen ions to opacity can not be inferred, as reported in this work. In future works we plan to compute and compare the total monochromatic and Rosseland Mean Opacity of oxygen under various temperature-density conditions, including those measured at the Sandia Z.

The present results are limited to monochromatic opacities for individual oxygen ions and do not include total plasma opacities, which also include free-free and photon scattering components arising from all ions present in the plasma, including those that are fully ionized. A careful treatment of these components is in progress and results will be presented elsewhere for a wider grid of temperatures and densities pertaining to stellar interiors.

The present work forms part of a broader investigation into oxygen and iron contributions to total plasma opacities, in particular under solar interior conditions, and full solar mixture opacity calculations with varying elemental abundances. The aim is to address discrepancies in standard stellar models using existing opacities. These ion-resolved data are also directly relevant to transmission-spectrum analysis in ICF and other HED plasma experiments.

% Each of the commands below will create an unnumbered section with the appropriate heading.
% Remove any sections that are not relevant for your article.
% All sections except suppdata will be removed if the [anonymous] option is used.
% See iopjournal-guidelines.pdf for more information.
%

\ack{The Computational work was carried out at the Ohio Supercomputer Center, and on the Unity cluster at the Ohio State University.}

\funding{This work was supported by the U.S. National Science Foundation from grant Ast-2407470.}

%\roles{Sample text inserted for demonstration.}

\data{
All atomic data will be available online at the NORAD-Atomic-Data database \cite{norad} at the Ohio State University at:
https://norad.astronomy.osu.edu/}. The data for opacities will be reported separately, but may be obtained from the authors upon request.
% For more information on IOP Publishing's research data policy see: https://publishingsupport.iopscience.iop.org/questions/research-data/

%\suppdata{Sample text inserted for demonstration.}

\pagebreak
\bibliography{references}

\end{document}